\documentclass[oldversion]{aa}  

\makeatletter
\let\linenumbers\relax
\let\runningpagewiselinenumbers\relax

\let\@LN@col\@gobble

\makeatother

\newcommand{\lens}{AC\,114}

\newcommand{\lenstool}{\textsc{Lenstool}}

\usepackage{graphicx}
\usepackage{natbib}
\usepackage[hidelinks]{hyperref}

\begin{document}
 
\title{SLICE -- Combining Strong Lensing and X-ray in AC\,114
Further Insights into the Merger Scenario}
   \titlerunning{AC\,114}
   \authorrunning{Limousin et~al.}
   \author{Marceau Limousin\inst{1}, Benjamin Beauchesne\inst{2,3}, 
Keren Sharon\inst{4}, Dominique Eckert\inst{5}, Guillaume Mahler\inst{6}, 
Johan Richard\inst{7}, 
David Lagattuta,
Gourav Khullar\inst{8}, 
Mathilde Jauzac\inst{2,3,9,10},
Mike Gladders\inst{11,12},
Marco Balboni\inst{13,14},
Fabio Gastaldello\inst{14},
Stefano Ettori\inst{15},
Catherine Cerny\inst{4},
Eric Jullo\inst{1}, 
Gavin Leroy\inst{2}
\& Nency Patel\inst{2,3}
      \thanks{Based on observations obtained with the \emph{James Webb Space Telescope} and
the \emph{Hubble Space Telescope}.}
       }
   \offprints{marceau.limousin@lam.fr}

   \institute{
$^1$ Aix Marseille Univ, CNRS, CNES, LAM, Marseille, France.\\
$^2$ Centre for Extragalactic Astronomy, Department of Physics, Durham University, South Road, Durham DH1 3LE, UK.\\
$^3$ Institute for Computational Cosmology, Department of Physics, Durham University, South Road, Durham DH1 3LE, UK.\\
$^4$ Department of Astronomy, University of Michigan 1085 South University Avenue Ann Arbor, MI 48109, USA.\\
$^5$ Department of Astronomy, University of Geneva Ch. d’Ecogia 16, CH-1290 Versoix, Switzerland.\\
$^6$ STAR Institute, Quartier Agora - All\'ee du six Ao\^ut, 19c B-4000 Li\`ege, Belgium.\\
$^7$ Univ Lyon, Univ Lyon1, ENS de Lyon, CNRS, Centre de Recherche Astrophysique de Lyon UMR5574, 69230 Saint-Genis-Laval,
France.\\
$^8$ Department of Astronomy, University of Washington, Physics-Astronomy Building, Box 351580, Seattle, WA 98195-1700, USA.\\
$^{9}$Astrophysics Research Centre, University of KwaZulu-Natal, Westville Campus, Durban 4041, South Africa.\\
$^{10}$School of Mathematics, Statistics \& Computer Science, University of KwaZulu-Natal, Westville Campus, Durban 4041, South Africa.\\
$^{11}$ Department of Astronomy and Astrophysics, University of Chicago, 5640 South Ellis Avenue, Chicago, IL 60637, USA.\\
$^{12}$ Kavli Institute for Cosmological Physics, University of Chicago, Chicago, IL 60637, USA.\\
$^{13}$ DIFA – Universit\`a di Bologna, Via P. Gobetti 93/2, I-40129 Bologna, Italy.\\
$^{14}$ INAF - IASF Milano, via A. Corti 12, 20133 Milano, Italy.\\
$^{15}$ INAF – Osservatorio di Astrofisica e Scienza dello Spazio di Bologna, Via P. Gobetti 93/3, 40129 Bologna, Italy.\\
              }

   
  \abstract
   {
AC\,114 is a historically significant galaxy cluster, being one of the first strong
lensing clusters detected from the ground in the early 1990s, prior to the launch of the 
\textit{Hubble Space Telescope} (HST).
Despite this early prominence, no detailed lensing analyses have been carried out for
more than fifteen years.
We here study this cluster using \textit{James Webb Space Telescope} (JWST) 
imaging obtained as part of the \textit{Strong LensIng and Cluster Evolution} 
(SLICE) program, complemented by archival HST and 
X-ray observations. JWST data reveal ten new multiply imaged systems and 
enable the identification of conjugate substructures in several of the sixteen systems,
significantly increasing the number of strong lensing constraints. Using these data, 
we construct a parametric mass model with \textsc{Lenstool} and extend it by explicitly 
incorporating the \textsc{Chandra} data in a combined strong lensing+X-ray fit, 
following the methodology recently introduced by Beauchesne et al. (2024).
Our best-fit model reproduces the multiple images with an RMS of 0.4$\arcsec$ while 
simultaneously matching the X-ray data. The dark matter distribution is unimodal 
and centered on the brightest cluster galaxy, with a large core radius of 
83$\pm$5\,kpc, consistent with values 
reported in other strong lensing clusters. 
The strong lensing constraints require the inclusion of an external shear component 
which position angle
points unambiguously towards a nearby ($\sim$\,1\,Mpc), well defined mass concentration at the same redshift in the 
North-West, for which we propose the naming AC\,114b.
The spatial coverage of the \textsc{XMM-Newton} data encompasses the whole structure, allowing us to probe
the X-ray properties of the companion cluster and the thermodynamics of \lens, providing further evidence for
a major merger, in line with previous signatures seen in \textsc{Chandra}, radio and optical spectroscopic data.
Our results shed new light on the merging scenario, revealing a major merger caught in a late
post-collisional phase, where \lens\, is the dominant system and \lens b has likely been stripped of its hot gas.
Our analysis highlights the power of combining strong lensing constraints with X-ray data 
to disentangle the dark matter and gas components and to investigate the dynamical processes driving cluster mergers.
Our lens model and associated products are available for download at the 
\href{https://data.lam.fr/sl-cluster-atlas/}{Strong Lensing Cluster Atlas Data Base}, 
which is hosted at Laboratoire d'Astrophysique de Marseille.
   }


   \maketitle
 
\vspace{-2cm}

\section{Introduction}

The late 1980s marked the era of the first discoveries of giant gravitational arcs in galaxy clusters,
made using ground-based telescopes of the two-meter class
\citep{lynds1,Soucail_1987,Lavery_1988,Lynds_1989}.
These findings opened a new chapter in observational astronomy.
Accompanied by theoretical insights, these early detections led to the conclusion that
\emph{"the lensing hypothesis should be taken seriously"} \citep{Grossman_1988}.
In the early 1990s, pioneering studies transformed these initial discoveries from scientific curiosities
into powerful cosmological tools, especially with the advent of the \emph{Hubble Space Telescope} (HST).

Galaxy cluster \lens\footnote{Also known as Abell~S1077.}, 
at redshift 0.317, is one such example.
Lensing features were first detected in ground-based images \citep{Smail_1991},
then with WFPC1 onboard the HST \citep{Smail_1995}, 
and with WFPC2 shortly afterward \citep{Priya_1998}.
Multiply imaged systems were reported, primarily by \citet{Priya_1998}, and spectroscopic redshifts were measured 
for some of them  
\citep{Campusano_2001,Lemoine_2003}, enabling early lensing analyses \citep[\emph{e.g.}][]{Sereno_2010}.
The strong magnification of this cluster has made it a valuable target for the study of 
distant galaxies \citep{Richard_2006}.

\lens\, has been identified in major recent Sunyaev-Zeldovich (SZ) surveys.
It is detected in the Planck PSZ2 catalogue \citep{Planck_2016}, where it is 
referred to as PSZ2G008.31-64.74.
This detection was confirmed at high significance in the SPTpol Extended Cluster 
Survey \citep{Bleem_2020}, under the designation SPT-CLJ2258-3447, with a mass estimate
M$_{500}$\,=\,7.23$^{+0.82}_{-1.01}$ 10$^{14}$ M$_{\sun}$.
\lens\, is also reported in the Atacama Cosmology Telescope catalog \citep{Aguena_2026}
with a mass estimate M$_{500}$\,=\,5.94$^{+1.15}_{-0.97}$ 10$^{14}$ M$_{\sun}$.
The discrepancy between experiments is not unexpected, and the impact of such differences on
the SZ-mass scaling relations is disucssed in \citet{Hilton_2021}.

\lens\ was first observed in X-rays with the \textsc{Chandra} 
telescope by \citet{DeFilippis_2004}.
While a sub-arcsecond offset between the X-ray peak and the brightest cluster galaxy (BCG) 
could suggest a relaxed cluster \citep{Rossetti_2016}, the
irregular morphology and a $\sim$10$\arcsec$ offset between the X-ray centroid and
the BCG rather indicate some merging activity. 
In particular, a soft X-ray
tail is observed, connecting the cluster center to the South-East.

This picture is confirmed by further X-ray observations with the \textsc{XMM-Newton} 
telescope, conducted
as part of the \textsc{CHEX-MATE} survey \citep{chexmate}, which provided valuable
constraints on the thermodynamical state and morphology of the system.
Morphological classifications indicate a disturbed
structure, and the cluster is ranked among the most disturbed systems in \textsc{CHEX-MATE} 
\citep{Campitiello_2022}.
This is significant, since \textsc{CHEX-MATE} clusters are Planck SZ selected and not biased towards 
merging clusters \citep{Rossetti_2017}.
In addition, the temperature profile shows a flat central 
shape and the absence of a strong cool-core, both consistent with a merging 
state \citep{Iqbal_2023,Rossetti_2024}.

At radio wavelengths, \textsc{ASKAP} observations \citep{Duchesne_2024} report 
the detection of
a double radio relic system, along 
with additional
residual emission at the cluster centre that has been proposed as a candidate radio halo.
More recently, \textsc{MeerKAT} observations \citep{Balboni_2025} clearly detect central diffuse radio
emission, well aligned with the thermal emission, thus confirming its classification as a 
radio halo,
which is a characteristic signature of some merging activity.

This dynamical activity is further supported by spectroscopic surveys
\citep{Proust_2015,Saviane_2023,Andrade_2024,Pizzuti_2025,Sereno_2025}, which report 
a velocity dispersion approaching 2000\,km\,s$^{-1}$.

The present work provides additional evidence for this dynamical state, based on 
both strong lensing (SL) and a re-analysis of the \textsc{XMM-Newton} data, and offers insights into the 
merging scenario.

Recently, \lens\, was observed by JWST as part of the SLICE survey 
(PID: 5594, PI: Mahler), using NIRCAM filters
F150W2 and F322W2, with an effective exposure time of 1836\,s in each filter.
The NIRCAM data reduction is described in \citet{Cerny_2025b}.
We also use archival HST data from program GO-11591, obtained from MAST and processed with 
\textsc{Drizzlepac/AstroDrizzle} to match the reduced JWST/NIRCAM imaging pixel scale and astrometry.
The dataset includes ACS/F814W (4920 + 2380\,s exposure) and WFC3/F160W and F110W (2412\,s each).

JWST data do 
reveal new multiply imaged systems and offer additional insights into the 
previously known ones.  
In this paper, we present these newly identified systems and construct a parametric mass model using 
the \lenstool\ software \citep{jullo07}.  
We then combine these data with \textsc{Chandra} X-ray 
observations to perform a strong lensing + X-ray (SL+X-ray) analysis of \lens, with the goal of probing the mass distribution of this historically significant galaxy cluster.

We present the multiple images in 
Section~2, which serve as constraints for the strong lensing mass model described in Section~3, 
where we detail the various steps involved in elaborating 
the mass model.  
We then turn to the available X-ray data for \lens\, in Section~4.
We describe how the gas mass is explicitly incorporated into the mass model in order 
to construct our best-fit mass model of \lens\, presented in Section~5. 
In addition, we re-analyse the \textsc{XMM} data in order to derive further thermodynamic 
quantities relevant
for probing the dynamical state of \lens, which is discussed in Section~6.

All our results adopt the $\Lambda$ Cold Dark Matter ($\Lambda$CDM) concordance cosmology, with
$\Omega_{\rm{M}} = 0.3, \Omega_\Lambda = 0.7$, and a Hubble constant
\textsc{H}$_0 = 70$ km\,s$^{-1}$ Mpc$^{-1}$.
At the redshift of \lens\, ($z$\,=\,0.317), this cosmology implies a scale of
4.63\,kpc/$\arcsec$.
In all figures, North is up and East is left.

\section{Multiple Images}

We begin by presenting the previously known multiple images, which we have renamed for consistency.  
As part of this process, we verified the reliability of these images using the JWST data.  
We then introduce the new systems identified in the JWST observations (Fig.~\ref{fig1}).  
All multiple images are located within a circle of radius 30$\arcsec$ (139\,kpc), centred on the 
BCG. 
They are listed in Table~\ref{multiple}.
We note that all the previously known multiple images are located in the North-West part of the 
cluster.

\subsection{Previously Identified Systems}

Six multiply imaged systems were reported prior to the JWST observations.  
They are marked with red circles in Fig.~\ref{fig1}, and their coordinates and redshifts are listed in Table~\ref{multiple}.

\begin{itemize}
\item System S was proposed by \citet{Smail_1995}, who reported two images, S1 and S2, both at a spectroscopic redshift of 1.86.  
Image S3 was later identified by \citet{Priya_1998}. 
The earlier redshift estimate was later confirmed by
\citet{Campusano_2001} and by \citet{Lemoine_2003} at $z=1.867$.
This corresponds to system 1 in the present work.

\item System A was proposed by \citet{Smail_1995}, who reported six images, one of which at a spectroscopic 
redshift of 0.639.  
\citet{Priya_1998} later identified five images belonging to the same system, with a predicted redshift of $1.67 \pm 0.15$ from their lens model, including a radial arc composed of two merging images.  
Based on a single emission line, \citet{Campusano_2001} measured a spectroscopic redshift of 1.691 for images A1 and A2, which was later revised to 1.869 by \citet{Lemoine_2003}.  
This corresponds to system 2 in the present work, following the identification proposed by \citet{Priya_1998}.
\item System B was proposed by \citet{Priya_1998}, who reported five images with a predicted redshift of $1.17 \pm 0.10$.  
\citet{Campusano_2001} detected only a single emission line in their spectroscopic observations and were unable to distinguish between a redshift of 1.5 and 2.1.  
This corresponds to system 3 in the present work, which redshift is optimised by the mass model.

\item System C was proposed by \citet{Priya_1998}, who reported three images of a "partial-ring shaped galaxy," with a predicted redshift of $2.1 \pm 0.30$.  
\citet{Campusano_2001} measured a spectroscopic redshift of 2.854 for image C3.
This corresponds to system 4 in the present work.
\item System D was proposed by \citet{Priya_1998}, who reported five images with a predicted redshift of $1.18 \pm 0.10$.  
This corresponds to system 5 in the present work, which redshift is optimised by the mass model.

\item System E was proposed by \citet{Campusano_2001}, who reported five images and measured a spectroscopic redshift of 3.347 for image E1.  
This corresponds to system 6 in the present work.
\end{itemize}

\subsection{New Systems}

\begin{figure*}
\begin{center}
\includegraphics[scale=0.4,angle=0.0]{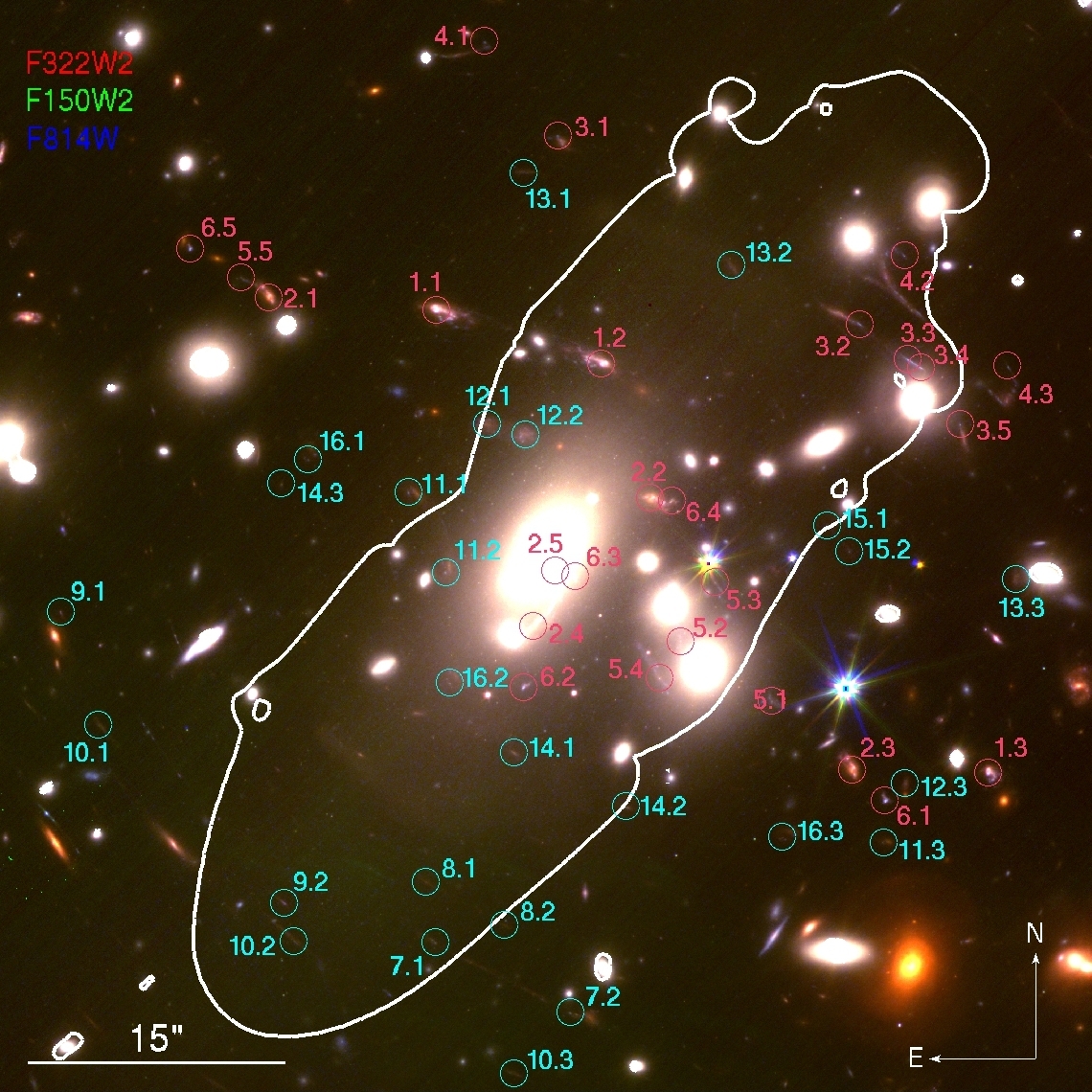}
\caption{Color image of the core of \lens\, produced with SAOImage ds9, using JWST/NIRCam F322W, F150W, and HST/ACS F814W in the red, green, and blue channels, respectively.
We show in red the multiple images known before \textsc{JWST}, and in cyan the one 
discovered thanks to the \textsc{JWST} data.
We draw in white the critical curve for $z=1.87$, the redshift of
systems 1 and 2.}
\label{fig1}
\end{center}
\end{figure*}

We identified ten new multiply imaged systems thanks to the JWST data (highlighted with 
cyan circles in Fig.~\ref{fig1}).  
Most of these systems are located in the southern part of the cluster, where no multiple images had been reported prior to the JWST observations.  
The identification of the 10 new multiply imaged systems is based on visual inspection of the JWST images by several pairs of eyes (namely ML, BB, KS, GM, JR, DL, MJ, CC and EJ).
As is usually done, a multiply imaged system should satisfy the following criteria: similar colors and a geometrical appearance compatible with a lensing configuration. This identification process is a step by step process using the predictive power of the lens model. We started with a first lens model constructed from the previously identified systems 
in order to test a new system we have repered by eyes. Then this system is given as an extra constraint to the current lens model. Then the optimisation of a new model using this new system helps to assess its reliability. 
More precisely, we verify that the RMS of the new model,
which includes this new system, remains stable, \emph{i.e.}
it does not change by more than $\sim$0.2$\arcsec$.
And so on for the 10 systems.
No spectroscopic redshift measurements are currently available. The global uncertainty on the 
lens model takes into account this unknown redshifts, which are let free during the 
optimisation, adopting a broad prior, ranging from 1 to 10. 
Estimates on these redshifts are given in Table~\ref{multiple}.

\subsection{Conjugating Sub-Spots in the Images}

Thanks to the high-resolution capabilities of JWST, we are able to identify and conjugate emission knots in several of the lensed galaxies—specifically those corresponding to systems 1, 2, 3, 4, 6, 7, and 11. This feature was already noted in the JWST-SLICE data \citep{Cerny_2025b}.
As a result, we increase the total number of images used as constraints from 52 to 124. This corresponds to an increase in the number of constraints from 72 to 166.  
It is worth noting that while conjugating sub-spots in multiple images provides additional constraints, they are not as strong as those obtained from newly identified multiply imaged systems.
We quantify the added value of these sub-spot constraints in Appendix~\ref{subspots_Appendix} by comparing the constraints on mass model parameters with and without their inclusion. 
Appendix~\ref{subspots_Appendix} also lists and illustrates the sub-spots that we have been able to conjugate. 
Our final multiple image catalog, used in all subsequent analyses, includes these sub-spots.

\section{Strong Lensing Modelling}

We describe here our parametric strong lensing (SL) mass modelling of \lens, based on the multiple-image systems presented in the previous section.  
All mass components are described using the dual pseudo-isothermal elliptical mass distribution (dPIE) profile. 
We refer the reader to \citet{Limousin_2005} and \citet{ardis2218} for a description of
this mass profile. Here we only give a brief overview.
The geometrical parameters are the position, ellipticity and position angle.
Then it is parametrised by a fiducial velocity dispersion, $\sigma$, a core radius,
$r_{\rm{core}}$, and a scale radius, $r_s$, usually fixed to 1500\,kpc for cluster scale DM 
haloes since SL cannot constrain it.
Between $r=0$ and $r=r_{\rm{core}}$, the mass density is constant.
Then between $r=r_{\rm{core}}$ and $r=r_s$, the mass density is isothermal ($r^{-2}$), then it falls as $r^{-4}$
beyond $r_s$.

The photometric catalog for identifying cluster members was generated 
using Source Extractor \citep{sextractor} in dual-image mode in F814W and 
F160W, with F814W used as the detection image. The magnitudes and magnitude 
errors were measured as MAG\_AUTO with DETECT\_THRESH=10 and DETECT\_MINAREA=5.
Cluster members were selected from a F814W\,-\,F160W vs F160W color-magnitude 
diagram using the red sequence technique \citep{Gladders_2000}.

We include all cluster members within 50$\arcsec$ of the centre of \lens,
down to $m = 25$ (HST F160W).
This radius is sufficient to encompass the strong lensing region,
which lies within 30$\arcsec$ of the cluster centre. Cluster members located
outside this area are not included, as they have a negligible impact on the positions of the
multiple images while unnecessarily increasing the computational time.
This results in a sample of 107 cluster members. Each galaxy was visually
inspected to confirm its likely cluster membership.
In addition, to assess the reliability of the red-sequence selection, we considered publicly available spectroscopically confirmed cluster members and verified whether they are recovered by the red-sequence method.
We searched the NASA/IPAC Extragalactic Database (NED\footnote{https://ned.ipac.caltech.edu})
for redshifts of objects located within a circular region of radius 2 arcmin centered on the BCG.
From this sample, we selected cluster members as galaxies with velocities within 
$\pm$2000\,km\,s$^{-1}$ of the BCG redshift.
Finally, we restricted the sample to galaxies located within 50$\arcsec$ of the BCG.
A visual inspection revealed that two objects did not have counterparts in the HST or JWST images at
the reported coordinates, hence not considered.
This resulted in a final sample of 21 spectroscopically confirmed cluster members, all of which are successfully selected by our red-sequence method.

We use the \lenstool\ software, and all optimisations are performed in the image plane.  

\subsection{First model}

We begin by describing the mass distribution in \lens\, as the superposition of one large scale
DM halo and galaxy-scale perturbers associated with cluster members, 
which are included via a scaling relation that links their mass to
their luminosity \citep[see, \emph{e.g.}][]{mypaperIII}.
We \emph{impose} the position of the cluster scale DM halo to coincide with the light distribution, 
within a dozen of kpc, 3$\arcsec$ in practice, consistent with the typical offset 
allowed in a self-interacting dark matter 
(SIDM) scenario \citep[see discussion in][]{Limousin_2022}.

Regarding the luminosities of the cluster members, we use magnitudes in the HST F160W band, 
allowing us to adopt the 
results from \citet{Bergamini_2019} to constrain the scaling relations and limit degeneracies between 
the smooth large scale component and the galaxy-scale components \citep[see discussion in][]{Limousin_2016}.
\citet{Bergamini_2019} propose a Gaussian prior on the velocity dispersion of a galaxy with 
magnitude $m=17.05$ (F160W) of $248 \pm 28$ km\,s$^{-1}$ for MACS\,0416. We impose a Gaussian 
prior of $248 \pm 50$ km\,s$^{-1}$, broadening the uncertainty to account for differences between 
the populations of cluster galaxies.

We find that this initial model reproduces the multiple images with an RMS 
precision of 0.66$\arcsec$, which is noteworthy given the simplicity of the parametrisation.

The optimised position of the DM clump coincides with the centre of the stellar component
and is not stuck to any bound of the adopted prior.
If we enlarge this prior beyond what might be allowed by SIDM, 
\emph{i.e.} setting it to $\pm$ 15$\arcsec$ from the centre of the stellar
component, the optimised position remains coincident with the centre of the stellar component,
and all model parameters are consistent with the model in which the prior on the position
of the DM is set to $\pm$ 3$\arcsec$ from the centre of the stellar component.
The same holds if we set this prior to $\pm$ 25$\arcsec$.
This demonstrates that the coincidence between the DM halo and the BCG is not driven by the adopted prior.

\subsection{Optimising the BCG Individually}
We then remove the BCG from the galaxy catalogue and optimise it individually. 
Indeed, the BCG is not representative of the cluster galaxy population.
Moreover, several multiple images lie close to the BCG, including the radial arc 
(images 2.4 and 2.5) as well as image 6.3. The overall RMS decreases to 0.49$\arcsec$,
with a notable improvement in the RMS of the images located near the BCG.

We find that the optimised position of the BCG coincides with the centre of the stellar component.
The ellipticity, position angle, and scale radius of the BCG are found to be
unconstrained, while its core radius converges to 0. 
We therefore adopt a circular halo in the following, with a vanishing core radius and a fixed scale
radius of 40\,kpc, a value that does not influence the results.

The ellipticity of the cluster scale DM component is high, with a value of 0.80$\pm$0.01, which is
larger than typically
expected for a unimodal cluster \citep{Despali_2016}.
Its position angle is 54$\pm$1 deg, in agreement with the position angle of the
light distribution of the BCG.
We note that such an alignment between the two position angles could 
suggest a relaxed rather than disturbed state. This will be further discussed in 
Section~6.4.
The core radius of the DM component is 83$\pm$5\,kpc.
The high ellipticity of the large scale DM component motivates us to examine the surroundings
of \lens, as such a pronounced ellipticity may indicate the presence of unaccounted mass 
in the outskirts \citep{keeton97}.

\subsection{Exploring the Environment of \lens}

\begin{figure*}
\begin{center}
\includegraphics[scale=0.89,angle=0.0]{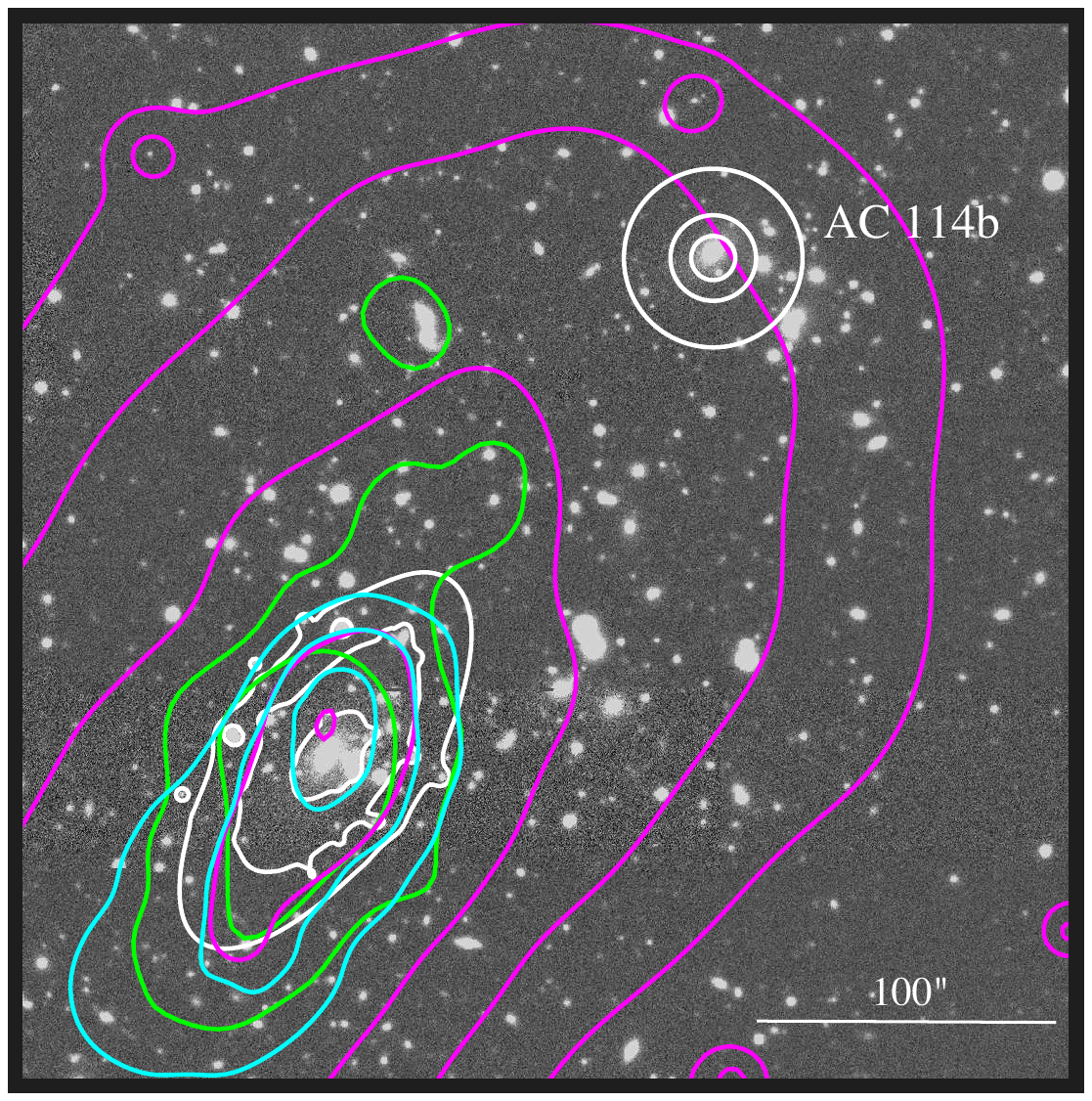}
\caption{
$i$ band image of the \lens-\lens b complex from Legacy Survey DR10 data.
In white, we show contours of the total mass as inferred from the SL analysis, where
\lens b is modeled as a singular isothermal sphere.
Contours of the X-ray count maps from \textsc{Chandra} and \textsc{XMM-Newton} are shown in cyan and magenta, respectively, while radio contours from \textsc{MeerKAT} are shown in green.
\lens b is dominated by
BCG$_2$, located at the centre of the northeast mass contours in this figure. 
It lies at a projected distance of 210$\arcsec$ (972\,kpc) from the BCG.
100$\arcsec$ corresponds to 463\,kpc.}
\label{largescale}
\end{center}
\end{figure*}

We examined the Legacy Survey\footnote{https://www.legacysurvey.org.} DR10 data around the 
position of \lens\, (Fig.~\ref{largescale}),
identifying an overdensity of bright elliptical galaxies to the North-West.
The brightest galaxy at (RA, Dec) = (344.6579, -34.7557) has a spectroscopic redshift of
0.317 \citep{Proust_2015}. We hereafter refer to this galaxy as BCG$_2$.
It is located 210$\arcsec$ (972\,kpc) from the centre of \lens, 
aligned with the position angle of the DM halo describing \lens.
A redshift of 0.317 is also measured for several bright elliptical galaxies located in 
this area \citep[see Fig.~1 in][]{Andrade_2024}.

Therefore, a companion cluster (\lens b hereafter) is located in the 
North-West of \lens\, and
must be taken into account in the modelling.
In fact, some elliptical galaxies are found between \lens\, and \lens b, suggesting the presence of a 
filamentary structure connecting the two components.
We then realized that \citet{Krick_2007}, who studied the intracluster light (ICL) in the field of \lens, 
had already mentionned \lens b, but only one redshift was
available at that time, preventing the authors from concluding wether \lens b was physically
associated with \lens.
They reported a centralised ICL component associated with \lens, as well as a diffuse component associated
with \lens b.

We include \lens b as an external shear component in the modelling.
This indeed slightly improves the fit, with the RMS dropping to 0.43$\arcsec$. Furthermore, 
the ellipticity of the main DM clump decreases to $\sim0.74$, which 
lies at the higher end of values expected from numerical 
simulations \citep{Despali_2016}.
The strength of the external shear is $\sim0.13$. 
The angle of this external shear\footnote{This corresponds to the angle of the shear generated
by the external mass component and experienced by the SL constraints.
Like other angles, it is defined with respect to the X axis (East-West direction), 
measured positively counter-clockwise.}, 
constrained to be equal to $\sim139^{\circ}$, points unambigously and consistently towards \lens b as the 
source of this shear.
We also include \lens b as a singular isothermal sphere centred on BCG$_2$.
The resulting RMS is 0.49$\arcsec$, comparable to the RMS obtained when using an external shear,
and the velocity dispersion of \lens b is equal to 
750$\pm$150\,km\,s$^{-1}$.
The ellipticity of the main DM clump decreases to $\sim0.74$.
In all models investigated hereafter (\emph{i.e.} the SL+X-ray models), \lens b is 
modelled as an external shear component, not as an SIS profile.
The SIS description is only used in Fig.~2 in order to present mass contours associated with 
\lens b and to illustrate the bimodality of the mass distribution in the 
\lens$-$\lens b system.

To summarize, our SL only analysis provides strong evidence for the presence of a massive companion cluster to \lens\,
in the north-west, which we propose to name \lens b.

Note that we do not attempt to further improve the RMS by optimising individual cluster members
located close to certain multiple images (namely images 2.4, 3.3, 3.4, 4.1, 4.2).

Finally, we have examined how the model responds to brighter magnitude cuts in the cluster 
member catalog.
When applying magnitude cuts at 24, 23, and 22, the RMS remains at 0.43$\arcsec$.
For a cut at magnitude 21, the RMS increases slightly to 0.44$\arcsec$. 
The model parameters remain consistent regardless of the magnitude cut applied.
This suggests that the model is primarily sensitive to the brightest cluster members.

\section{X-ray data sets}

\subsection{Presentation}

\textsc{Chandra} observations of \lens\,\citep{DeFilippis_2004} total
74.3\,ks of exposure time.
The observations are available in the \textsc{Chandra} Data Collection 
(CDC) $453$\footnote{~\href{https://doi.org/10.25574/cdc.453}{doi:10.25574/cdc.453}.}. 
The data have been reduced and analyzed following the procedure detailed in 
\citet{Beauchesne_2024} with \textsc{CIAO}\footnote{https://cxc.cfa.harvard.edu/ciao/} $4.17$ \citep{ciao} and \textsc{CALDB} $4.12.0$. To measure the cluster temperature, 
we limit our analysis to the $[0.5,7.5]$~${\rm keV}$ energy range and we use a hydrogen 
column density $n_{\rm H}$ of $1.079\times10^{20}$~${\rm atoms/cm^2}$. We obtain $n_{\rm H}$ 
with $n_{\rm H\,I}$ measurement from \citet{HI4PI} for which we account for molecular 
hydrogen following \citet{Willingale2013}.

\lens\, was observed by \textsc{XMM-Newton} on June 11, 2018 for a total of 45\,ks (observation ID 0827010901). 
We processed the observation using the XMMSAS v22.0 package and the X-COP analysis 
pipeline \citep{Ghirardini_2019}. After running the standard event screening procedures, we extracted light 
curves of the observation from the three cameras of the European Photon Imaging Camera (EPIC) and filtered out 
the time periods affected by flaring background. The clean observing time after flare filtering is 29\,ks (PN), 
35\,ks (MOS1), and 37\,ks (MOS2). From the clean event lists, we used the \textsc{mosspectra} and 
\textsc{pnspectra} executables to extract count images, exposure maps, and non X-ray background maps of the 
full field of view in the [0.7-1.2] keV band, which maximizes the signal-to-background ratio. We then combined 
the data from the three cameras to generate a single EPIC image. 

To extract X-ray emission contours, we used \textsc{asmooth} \citep{Ebeling_2006} to smooth the EPIC image 
and generate an adaptively smoothed, exposure corrected, and background subtracted map. We then extracted 
isophotes from the resulting map. For more details on the data processing technique, we refer the reader 
to \citet{Rossetti_2024}. 

\subsection{\textsc{Chandra} data: combining SL with X-ray}

Fig.~\ref{xray} shows the map of the X-ray counts, with the \textsc{csmooth}\footnote{https://cxc.cfa.harvard.edu/ciao/ahelp/csmooth.html}
\citep{Ebeling} contours overlaid in cyan.
The morphology is irregular, typical of a cluster undergoing dynamical
activity, with a soft X-ray tail extending toward the South-East. 

We here use the \textsc{Chandra} data, as it offers the highest spatial resolution among X-ray observatories while fully covering the strong lensing region.
Our goal is to explicitly include the X-ray gas mass in a combined strong lensing+X-ray (SL+X-ray) fit, following the methodology recently proposed by \citet{Beauchesne_2024} and implemented in \textsc{Lenstool}.
We refer the reader to that work for a detailed description of the modeling procedure.
To model the ICM mass, we rely on its X-ray emission,
using a combination of surface brightness and spectral data.
The surface brightness is derived from photon count maps,
while the spectral information provides the conversion factor needed to relate surface
brightness to gas mass.

\subsubsection{Fitting the X-ray data only}

We begin by fitting the X-ray data only with \lenstool. 
In this analysis, we consider the map of X-ray photon counts along with the relevant 
associated inputs.
Our goal is to describe the distribution of the X-ray gas mass using a superposition of dPIE 
profiles, which will later be incorporated into the combined SL+X-ray fit.

We first investigate how many dPIE clumps are required to adequately reproduce the X-ray only data.
As discussed in \citet{Beauchesne_2024}, we consider a model of the X-ray gas to be reliable if the likelihood of the observed data lies within the 5$\sigma$ credible interval (CI) of the expected likelihood for that model.

\begin{figure*}
\begin{center}
\includegraphics[scale=0.48,angle=0.0]{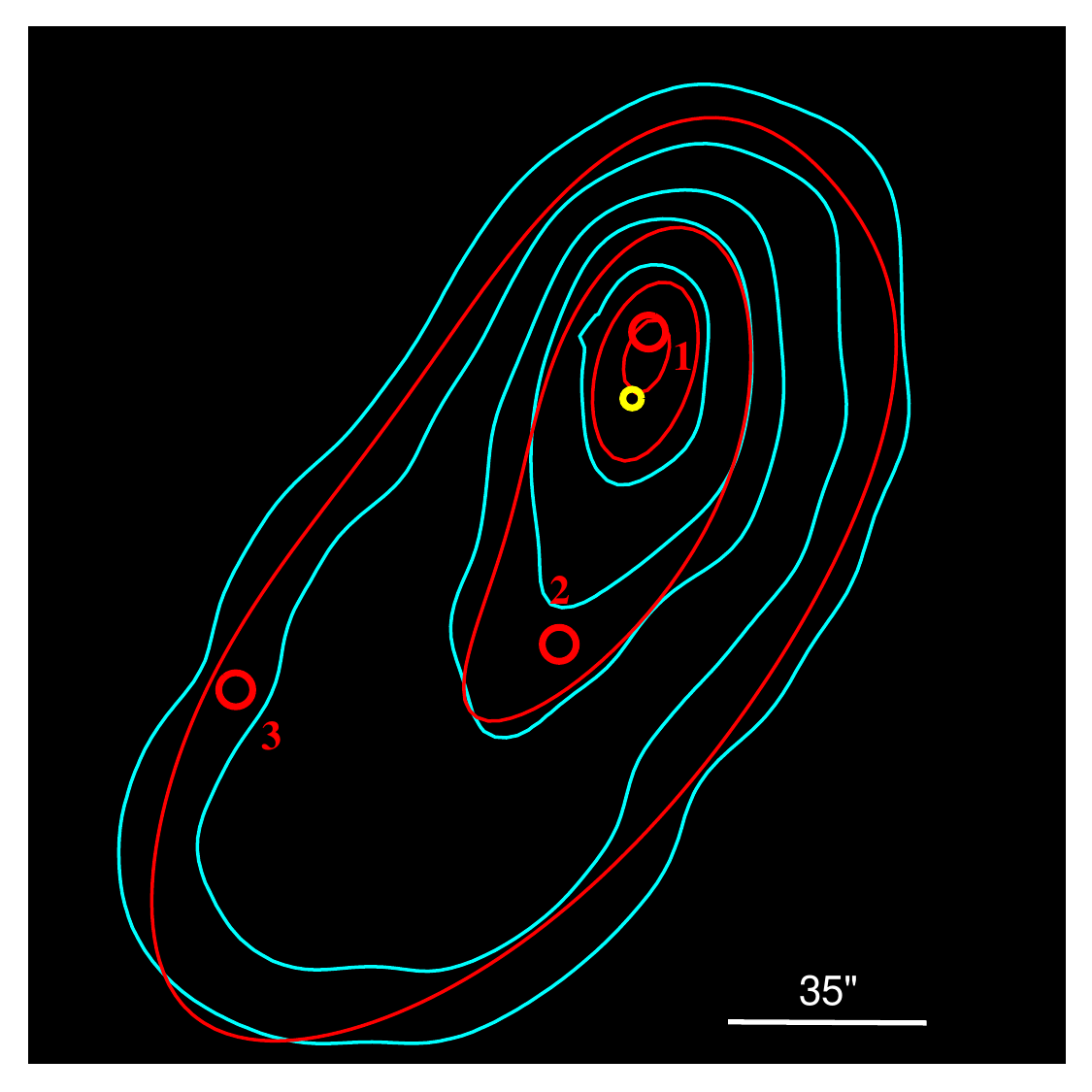}
\includegraphics[scale=0.48,angle=0.0]{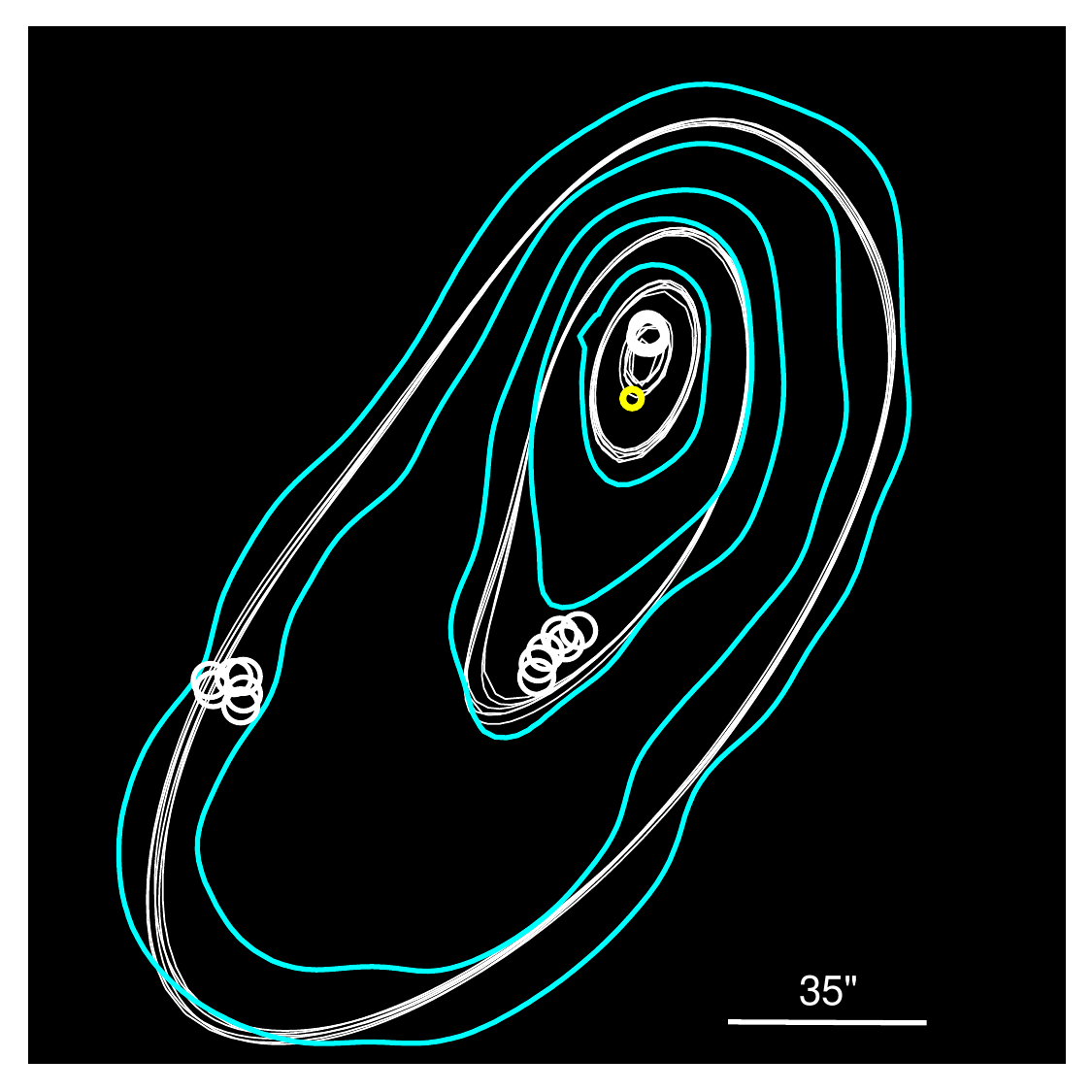}
\caption{\textsc{csmooth} \textsc{Chandra} X-ray contours are shown in cyan. 
The position of the BCG is marked by a yellow circle.
\emph{Left:}
Mass contours derived from fitting the X-ray data \emph{only} are shown in red.
The gas distribution is modeled as a superposition of three dPIE profiles, whose positions
are indicated by red circles.
\emph{Right:}
Mass contours, shown in white, correspond to different models of the gas distribution used 
in the combined SL+X-ray fits.
The positions of the dPIE components for each realisation are indicated by white circles.}
\label{xray}
\end{center}
\end{figure*}

Following \citet{DeFilippis_2004}, we begin by modeling the X-ray data as a superposition 
of two components: one located in the BCG region, and the other in the Southern part of 
the field. However, this model is not considered reliable, as the resulting 
likelihood does not lie within the 5$\sigma$ CI.
We then introduce a third dPIE clump, allowing its position to vary freely within 
$\pm70\arcsec$ of the cluster centre. This addition improves the fit, with the resulting 
likelihood falling within the 3$\sigma$ CI. 
Including a fourth dPIE clump still provides an acceptable description, as the likelihood 
remains within the 5$\sigma$ CI, although it lies outside the 3$\sigma$ CI.
We therefore adopt the simplest model composed of three dPIE components to describe the mass 
associated with the X-ray emitting gas. In Fig.~\ref{xray} (left panel), we show the positions of these 
clumps and the resulting mass map, along with the \textsc{csmoothed}
contours of the X-ray photon counts.

Note that the dPIE clumps used to describe the X-ray gas mass should not be 
interpreted as "regular" mass clumps. 
Their parameters are not representative of
physical mass clumps like the ones used to model the DM components, 
nor of clumps that typically arise in numerical simulations.
This approach is similar to the method presented in \citet{Jullo_2009}, where a grid 
of dPIE components is used to describe mass distributions in a kind of "free-form" 
modeling \citep[see also][]{my0854}.

\subsubsection{Combined Strong Lensing \& X-ray fit}

We now turn to a combined SL+X-ray fit. 
Technically, we merge the parameter file used in the SL only analysis with 
the one used in the X-ray only analysis.
This results in a model where the following mass components are optimised: 
the dark matter (DM) clump, the BCG, the external shear component, 
the galaxy scale component,
and the three dPIE profiles describing the X-ray gas.
In practice, we perform six identical SL+X-ray fits to assess the stability 
of the results. 
In particular, we aim to determine whether the addition of the three dPIE mass 
components introduces further degeneracies.
These six combined models are found to be stable, in the sense that the RMS values, 
the description of the gas component, and the remaining model parameters are all
consistent across the different runs. 
We note that six is an arbitrary number; we consider that if six 
realisations of the same SL+X-ray model are in agreement, there is no need to add more.
We find that these models reproduce the multiple images as accurately as the SL only 
model, with RMS values ranging between 0.42$\arcsec$ and 0.44$\arcsec$.
In Fig.~\ref{xray} (right panel), we show the mass contours corresponding to the different prescriptions for the X-ray gas mass: they are in very good agreement with one another, as well as 
with the mass contours obtained from the X-ray only fit.
The positions of the dPIE components for each realisation are also found to be stable.

\subsection{\textsc{XMM-Newton} data: thermodynamic quantitites}

Fig.~\ref{largescale} shows the contours of the \textsc{XMM-Newton} X-ray counts map overlaid in magenta 
on the Legacy Survey image.
Comparison with the \textsc{Chandra} contours reveals a very good agreement between the datasets, 
in terms of the irregular morphology and the offset between the X-ray emission peak and its centroid. 
The \textsc{XMM-Newton} data are much more extended, reaching up to 2\,Mpc from the cluster centre, 
therefore encompassing \lens b. 
We can immediately note the absence of X-ray emission from \lens b, suggesting that 
any X-ray emitting gas that might have been associated with \lens b has been stripped.

The sensitivity of \textsc{XMM-Newton} allows us to derive important thermodynamical
quantities.
\citet{Rossetti_2024} already presented the temperature profile. 
We here revisit the \textsc{XMM} data in order to derive
the entropy profile, which is presented in the following. 

\section{Results}
\label{bestfit}

We present the description of \lens\, resulting from our combined SL+X-ray fit, as 
detailed in the previous sections.
We then present the results of the gas properties inferred from the \textsc{XMM-Newton} data.

\subsection{Best SL+X-ray Fit Model}

This model succeeds in reproducing both multiple images positions, with an RMS of 0.43$\arcsec$, 
and the X-ray data.
It was obtained using a \textsc{Rate} parameter of 0.005 and an \textsc{Nb} value of 2000. The (\textsc{Rate}, \textsc{Nb}) test is presented in Appendix~\ref{RateNbAppendix}. For a detailed discussion of these parameters and the associated testing procedure, we refer the reader to \citet{Limousin_2025}.

\begin{figure*}
\begin{center}
\includegraphics[scale=0.5,angle=0.0]{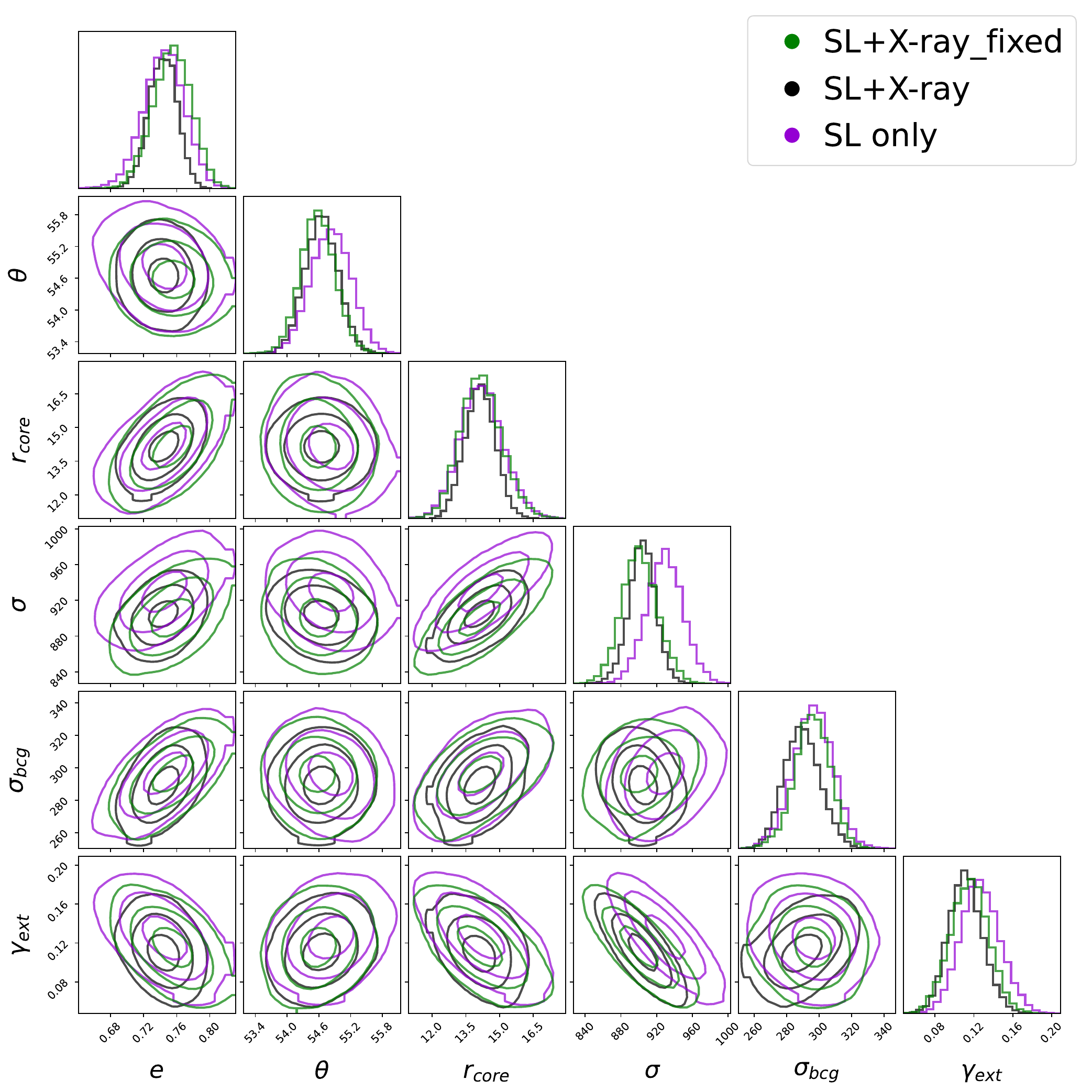}
\caption{Corner plots showing the posterior distributions of the main cluster
parameters for the SL+X-ray fit (black) and for the SL only fit (violet):
ellipticity $e$, position angle $\theta$, core radius $r_{\rm core}$ and
velocity dispersion $\sigma$ of the DM mass distribution; velocity dispersion of the BCG
$\sigma_{\rm bcg}$ and strength of the external shear $\gamma_{\rm ext}$.
We also show in green the results obtained by
fitting the SL constraints while including the description of the X-ray 
component
derived in Section~4.2.1 from the X-ray only fit.
The RMS is equal to 0.43$\arcsec$ in all cases.}
\label{compare_SL_only}
\end{center}
\end{figure*}

All parameters of the model — including the DM clump, the BCG, the galaxy-scale perturbers, the external shear, and the three dPIE profiles 
describing the X-ray gas mass component - are listed in Table~\ref{sl_xray_table}.
The PDFs of the \emph{main cluster parameters} (parameters of the DM clump, velocity dispersion of the
BCG, strength of the external shear) are shown in black in Fig.~\ref{compare_SL_only}.
Note that, for clarity, we do not display all model parameters in Fig.~\ref{compare_SL_only} 
nor in the other corner plots presented in this paper. We omit the parameters of the galaxy-scale perturbers,
and we do not show the posterior distributions of the position of the DM clump, as they always fall within the adopted prior of $\pm 3\arcsec$ from the centre of the BCG.
All MCMC chains are made publicly available, allowing anyone to investigate their behaviour if necessary.

Results of the best fit model are compared
with those derived using strong lensing (SL) constraints only 
(obtained in Section~3.3 and shown in violet in Fig.~\ref{compare_SL_only}).
The parameter PDFs are generally narrower when X-ray data are included, indicating improved constraints on the model parameters.
Parameters of the DM clump are consistent between both runs,
except for the velocity dispersion which is found to be lower when X-ray data are
included, as expected, since part of the mass is then attributed to the X-ray gas.
The strength of the external shear is also found to be slightly smaller when including the X-ray 
constraints. This is consistent with the shape of the gas mass component (Fig.~\ref{largescale}), which is 
elongated along the South-East North-West direction, \emph{i.e.} colinear with the
line connecting \lens\, and \lens b.
We also show in green in Fig.~\ref{compare_SL_only} the results obtained by 
fitting the SL constraints while including the description of the X-ray 
component 
derived in Section~4.2.1 from the X-ray only fit.
This approach is similar to the one proposed by \citet{Bonamigo_2018}.
The results are very close to those obtained from the full SL+X-ray analysis,
except for the strength of the external shear, which lies between the values
derived from the SL only and SL+X-ray models.
Moreover, the parameter uncertainty are smaller in the full SL+X-ray analysis,
highlighting the advantages of a fully combined analysis compared to a two step
approach.

\begin{table*}
\begin{center}
\begin{tabular}{cccccccc}
\hline \\*[-1mm]
Model & $\Delta$\,\textsc{ra} & $\Delta$\,\textsc{dec} &  $e$  & $\theta$ & $\sigma$  & $r_{\rm core}$ & $r_s$ \\
            &  $\arcsec$            &  $\arcsec$             &       &          &  km\,s$^{-1}$ & $\arcsec$ & $\arcsec$ \\
\hline \\*[-1mm]
DM   &  0.32$\pm$0.21  & -1.13$\pm$0.27  & 0.74$\pm$0.02  & 54.6$\pm$0.3  &  904$\pm$15 & 14.1$\pm$0.7 & [325]  \\   
        & (0.37)              & (-1.0)                 &   (0.72)              &  (54.9)              &       (895)      &  (14.0)  &  \\
\hline \\*[-1mm]
BCG   &  0.59$\pm$0.13  & -0.13$\pm$0.17  & --   &  --   & 289$\pm$11  &  [0] & [10.0] \\ 
        & (0.37)              &           (-0.06)       &      &       &  (284)             &     &        \\
\hline \\*[-1mm]
Gas \#1  &  3.15$\pm$0.38  & 13.24$\pm$0.67   & 0.51$\pm$0.03   &  68.3$\pm$1.8  & 168$\pm$5   &  18.1$\pm$0.8  & [270] \\ 
	& (2.69)                    & (12.69)           & (0.50)                    &    (68.7)            &    (172)            & (18.7)       &        \\
\hline \\*[-1mm]
Gas \#2  &  -12.5$\pm$1.9  & -41.7$\pm$2.3  & 0.81$\pm$0.02  &  45.8$\pm$1.2  & 272$\pm$12  &  62.4$\pm$3.8 & [270] \\ 
	& (-9.1)                    & (-39.0)           & (0.82)                    &    (46.3)            &    (270)            & (64.0)       &        \\
\hline \\*[-1mm]
Gas \#3  &  -68.5$\pm$1.5 & -49.9$\pm$3.6  & 0.20$\pm$0.09  &  126.5$\pm$11.3  & 328$\pm$10  &  110.7$\pm$5.7  & [270] \\ 
	& (-68.7)                    & (-52.3)           & (0.25)                    &    (127.8)            &    (325)            & (325.5)       &        \\
\hline \\*[-1mm]
\end{tabular}
\end{center}
\caption{dPIE parameters inferred for the mass model of \lens, constrained using both SL and X-ray
constraints, with an RMS equal to 0.43$\arcsec$.
Error bars on all parameters quoted in this paper correspond to 
$\pm$\,1$\sigma$ confidence levels inferred from the MCMC optimisation.
Coordinates are given in arcseconds relative to $\alpha$\,=\,344.701518, $\delta$\,=\,-34.802364;
$e$ and $\theta$ are the ellipticity and position angle of the mass distribution.
Each parameter is given as the median with the best fit value in parenthesis except the ones
fixed a priori that are shown in brackets.
For an L* galaxy, we have $\sigma$\,=\,187$\pm$5 km\,s$^{-1}$ and $r_s$\,=\,11$\pm$1.2$\arcsec$.
The external shear component is equal to 0.11$\pm$0.01, with an angle of 138.1$\pm$0.8.}
\label{sl_xray_table}
\end{table*}

\subsection{Mass Profiles of the different components}

\begin{figure*}
\begin{center}
\includegraphics[scale=0.5,angle=0.0]{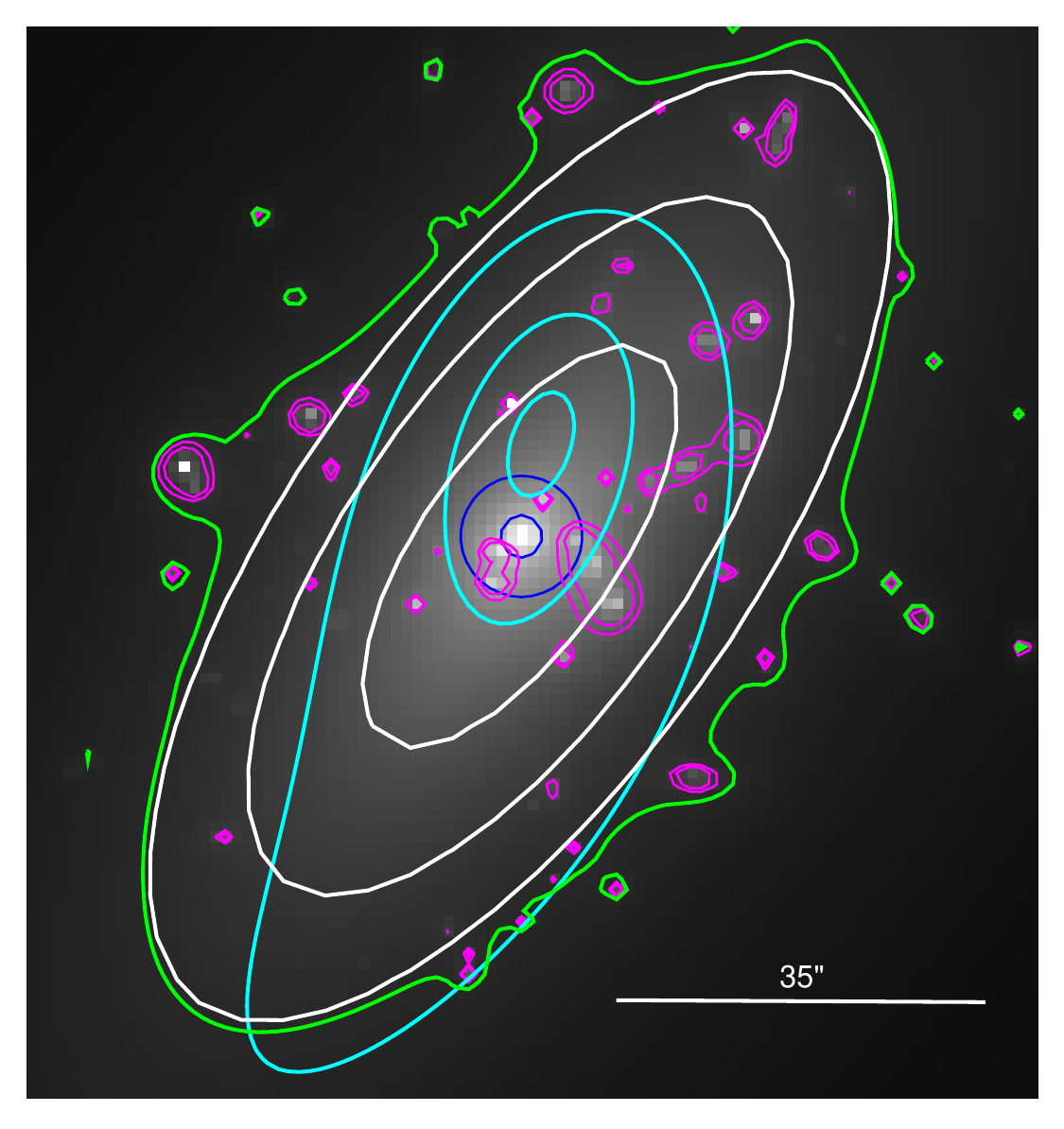}
\includegraphics[scale=0.5,angle=0.0]{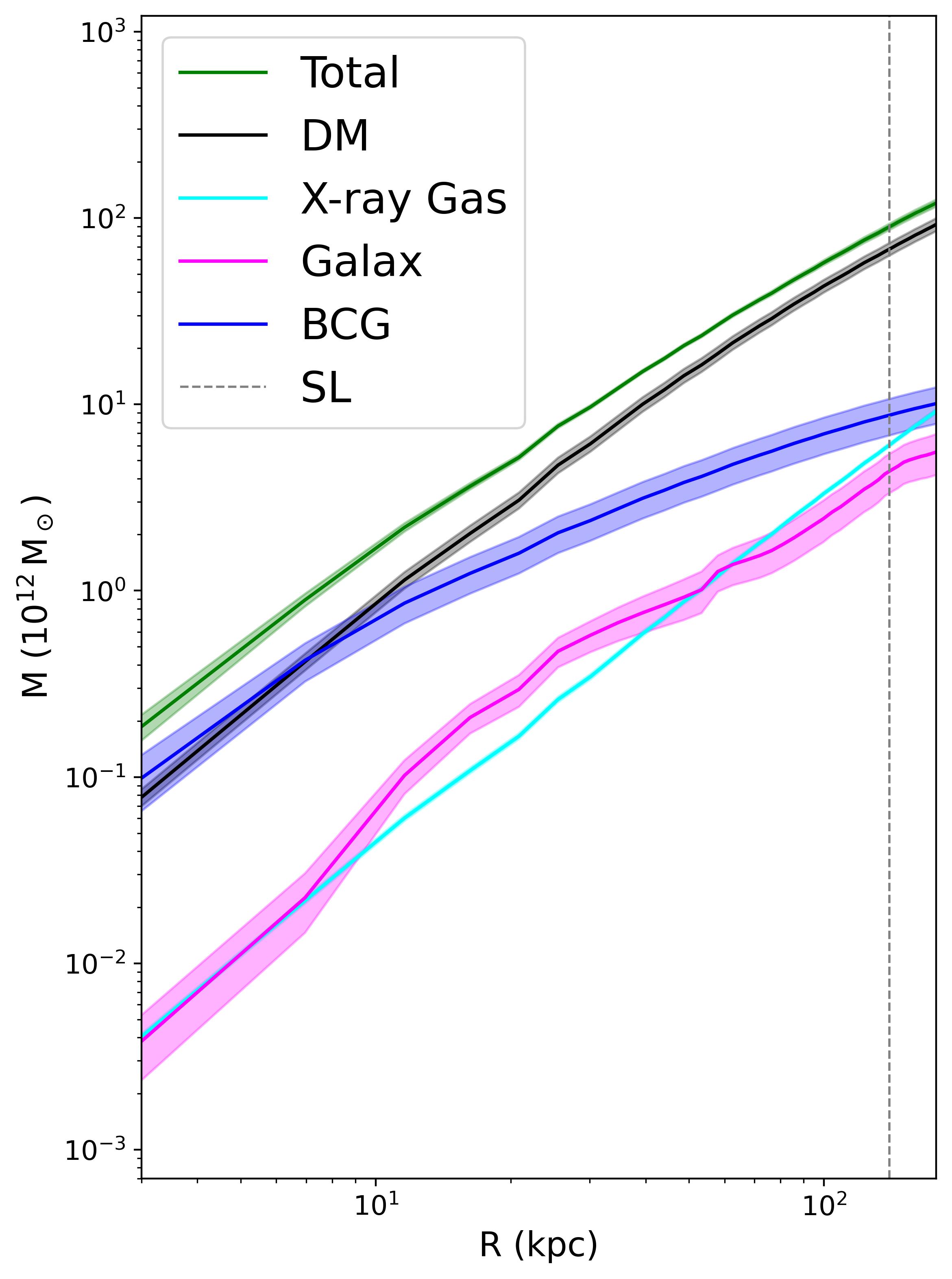}
\caption{\emph{Left:} 2D projected total mass map. 
White contours correspond to the DM component, cyan to the X-ray gas, blue to the BCG, and magenta
to the galaxy scale perturbers.
The green contour corresponds to the total mass. For clarity, we show only one.
Its similarity with the white contour highlights
that the DM component dominates the overall shape of the total mass map.
\emph{Right:} Corresponding 1D mass profiles for the total mass (green), the X-ray gas mass (cyan), the dark matter component (black), the BCG (blue), and the galaxy-scale perturbers (magenta).
The vertical dashed line corresponds to the extent of the SL region.}
\label{mass_profiles}
\end{center}
\end{figure*}

We show in Fig.~\ref{mass_profiles} the mass profiles of the different components describing \lens.
On the left, we present the 2D projected total mass map, overlaid with contours:
white for the DM component, cyan for the X-ray gas, blue for the BCG, and magenta
for the galaxy scale perturbers.
The green contour corresponds to the total mass. For clarity, we show only one.
Its similarity with the white contour highlights
that the DM component dominates the overall shape of the total mass map.

The shapes of the DM and gas mass distributions are similarly elongated along the 
South-East North-West direction, 
with their position angles differing by $\sim$18$^\circ$.

Regarding the positions of the different components and their offsets, we find that the peak of the mass associated with the BCG coincides with the centre of its light distribution. This position is also coincident with the peak of the mass associated with the main DM halo, as well as with the peak of the total mass inferred from strong lensing. However, we find an offset of $\sim 9\arcsec$ between this position and the centroid of the X-ray–emitting gas, as reported by \citet{DeFilippis_2004}.

Note that Fig.~\ref{mass_profiles} is restricted to the core of \lens\, and 
does not encompass the full field of view of interest when considering the major merger 
scenario involving \lens b. Nevertheless, the mass map of Fig.~\ref{mass_profiles} is highly 
elliptical, with a position angle pointing towards \lens b, which can be interpreted as 
a remnant of the major merger. 
To illustrate the bimodality of the whole structure, we consider the 
SL model in which AC114b is described as an SIS mass component (Section~3.3) and present 
the resulting 2D projected total mass contours in Fig.~2.
The bimodality can be appreciated, with one mass concentration associated with \lens\, 
and another associated with \lens b.

We integrate the mass maps in circular apertures starting from the centre of the BCG 
in order to derive the correponding 
1D mass profiles, which are shown in the right-hand panel of Fig.~\ref{mass_profiles}. 
The (1$\sigma$) error bars are derived from the MCMC realisations.
The vertical dashed line marks the radial extent of the strong lensing constraints.

\subsection{The Central Entropy} 
The central entropy characterizes the thermodynamical state of the cluster core and determines whether the system hosts a cool core \citep{Cavagnolo_2009}.
We estimated the central entropy, $K_0$, by fitting the deprojected entropy profile (Section~4.3) with a power law including a central floor \citep{Cavagnolo_2009}. We measure $K_0 = 131 \pm 26$ keV\,cm$^2$, indicating that the gas in the central regions has high entropy and classifying the system as a non–cool-core cluster \citep{Hudson_2010}.

\subsection{A Cored DM Distribution}
We have seen that the mass distribution of \lens, as described by a dPIE profile, features a 
core radius of 18$\pm 1 \arcsec$, which corresponds to 83$\pm 5$\,kpc.
Following the approach presented in previous works \citep{Limousin_2016,Limousin_2022,Cerny_2025a},
we repeat the analysis, imposing a non cored DM profile, \emph{i.e.}, forcing the core radius of the
dPIE profile to be smaller than 10\,kpc.
We obtain an RMS of 2.0$\arcsec$, and the X-ray likelihood lies outside the 5$\sigma$ credible 
interval. 
In order to compensate the need for a cored DM profile, 
the ellipticity of the DM halo increases to $0.83\pm 0.005$, and the strength of the external shear
is stuck to the upper bound of the prior (0.2), which is an unrealistically high value.
When this parameter is constrained to be smaller than 0.15, which is already at the upper end of what might be
expected, the RMS increases to 2.16$\arcsec$.

We therefore conclude that a cored DM distribution is favoured in \lens.
Such core-like
DM profile in strong lensing galaxy clusters have been reported for more than
two decades in many systems 
\citep[see, \emph{e.g.}][]{sand04,newman,Newman_2013,Bergamini_2019,Limousin_2022,Lagattuta_2023,Beauchesne_2024,Cerny_2025a}.
Some of these clusters display radial arcs, as \lens\, does, which are known to
form in shallow potentials \citep{Bartelmann_1996,Molikawa_2001}.
Large core radii are naturally explained in SIDM 
\citep[see][for a discussion]{Limousin_2022}. 
In the case of \lens, the core radius is comparable to that found in AS\,1063
using the same dPIE mass profile
\citep[$\sim 90$\,kpc, see][]{Bergamini_2019,Granata_2022,Limousin_2022}. 
Recent study by \citet{Diego_2026} interpreted the value of this 
core within SIDM, finding a self interaction cross section equal to 0.3\,cm$^2$\,g$^{-1}$.
If we interpret our result for \lens\, within SIDM, we therefore obtain the same
constraint on the self interaction cross section.

However, \citet{mypaperV} report a cusp-like DM profile in Abell~1703 using
SL techniques, suggesting a possible \emph{diversity} in the inner DM profiles of
strong lensing clusters.

\section{A Major Merger in a Post-Collisional Phase}

We argue that we are witnessing a major merger between \lens\, and \lens b in a post-collisional phase.
Evidence for this scenario is already present in the literature, and we provide further compelling evidence in support of this interpretation.

We use the term major merger to refer to a cluster–cluster interaction with a mass ratio of order ~1:1–1:3, as commonly adopted in merger‑rate and cluster dynamical studies 
\citep[see, \emph{e.g.}][]{Fakhouri_2020}.

\subsection{Radio evidence of a major merger in a post-collisional phase}
\citet{Duchesne_2024} report the detection of three radio sources.
Two of these are radio relics, located in the south-east and north-west regions.
The third source is classified as a candidate radio halo.

\citet{Balboni_2025} clearly detect central diffuse radio emission. 
This allows a robust classification of the emission as a radio halo.
They also confirm the south-east relic, which exhibits a clear spectral steepening towards the cluster centre.
In contrast, they argue that the north-west relic is more likely associated with a complex of radio-galaxy tails rather than a genuine relic.
We are therefore left with the firm detection of a radio halo and a radio relic.
The radio halo is shown in Fig.~\ref{largescale}, 
while the radio relic is located further to the southeast and is not shown here.

The correlation between a disturbed dynamical state and the presence of a radio halo has been firmly established since the early 
study of \citet{Buote_2001}. 
See also \citet{Cassano_2010}.
The widely accepted current scenario for radio-halo formation invokes the reacceleration of particles by turbulence induced during a merger
\citep[see the reviews by][]{Brunetti_Jones_2014,vanWeeren_2019}.

The radio halo is relatively bright even at 1.3\,GHz.
This indicates that a highly energetic merger has occurred, releasing a substantial amount of energy to power the diffuse radio emission.
The radio and X-ray emissions are well aligned (see Fig.~\ref{largescale}).
This suggests that the mechanism responsible for the radio-halo emission has also significantly modified the thermal ICM distribution.

Similarly, the location of the radio relic is well aligned with the south-east–north-west elongation of the system.
This suggests that the merger event occurred along this axis.
Indeed, radio relics are thought to trace shock waves associated with cluster mergers.
This is supported by simulations of cluster mergers 
\citep[\emph{e.g.}][]{Ricker_Sarazin_2001}.

Regarding the stage of the merger, the presence of both a radio halo and a relic suggests that the system is in a post-merger phase.

Radio halos require roughly 1\,Gyr to form \citep[e.g.,][]{Donnert_2018}. 
This is because the turbulent reacceleration mechanism that produces the diffuse emission is inefficient.

For the radio relic, the shocks that generate relics originate after the passage of the dark-matter cluster cores \citep[see][and references therein]{vanWeeren_2019}. 
However, the relic emission becomes visible only once these shocks reach the cluster outskirts \citep{Vazza_2012}. 
The typical shock crossing time is of the order of $\sim$1\,Gyr.

Numerical simulations show that the production of cosmic-ray electrons peaks about 1\,Gyr after core passage. 
These electrons eventually produce the radio synchrotron emission. 
The peak of the relic emission itself occurs slightly earlier, at roughly 0.7\,Gyr \citep{vanWeeren_2019, Nuza_2025}.

We conclude that radio observations provide evidence for a major merger in a post-collisional phase.

\subsection{X-ray evidence of a major merger}

X-ray morphological indicators have long been recognized as powerful probes of the dynamical state 
of galaxy clusters (see Buote et al. 2001 for an early review and 
Rasia et al. 2018 for a more recent one).
Dynamically relaxed systems exhibit regular X-ray isophotes and a centrally peaked surface-brightness profile, which are absent in dynamically disturbed clusters. In particular, post–core-passage mergers produce irregular X-ray morphologies and disrupt the central cool core.
In the case of \lens,
X-ray data from both \textsc{Chandra} and \textsc{XMM-Newton} reveal an irregular
morphology, a $\sim$10$\arcsec$ offset between the X-ray centroid and the BCG,
and a soft X-ray tail (Fig.~\ref{largescale}).

The joint use of multiple morphological parameters provides a powerful discrimination of the 
dynamical state of galaxy clusters, justifying the adoption of aggregated metrics such as 
the M parameter proposed by Campitiello et al. (2022; see Rasia et al. 2012 for earlier support). 
According to X-ray morphological indicators, \lens\, is dynamically active and ranks among the 
most disturbed clusters in the \textsc{CHEX-MATE} sample, with a disturbance rank of 100 out of 118.

Thermodynamical indicators for \lens\, further support the presence of a major merger. The temperature 
profile is flat \citep{Rossetti_2024} and does not display any cool core. Such a lack of a 
central cool core is generally characterized by elevated central entropy, consistent with our 
re-analysis of the \textsc{XMM-Newton} data (Section~5.3).

Morphological and thermodynamic indicators do not always provide consistent classifications 
of the dynamical state of galaxy clusters. For instance, in the Bullet Cluster (G266.04–21.25) 
the M parameter is low \citep[M=0.02][]{Campitiello_2022}, whereas spectroscopically derived 
thermodynamic quantities rank among the most extreme in a \textsc{CHEX-MATE} subsample 
\citep{Lovisari_2024}, clearly signaling a major merger.
When both classes of indicators agree, as in the case discussed here, this concordance constitutes 
additional, robust evidence that the system is a merger \citep[see][]{Hudson_2010}.

We conclude that X-ray observations provide evidence for a major merger in \lens.

\subsection{Dynamical Considerations}

In addition to the velocity dispersion approaching 2000\,km\,s$^{-1}$,
\citet{Pizzuti_2025} compute the Anderson–Darling statistic, A$^2$, which quantifies deviations 
from Gaussianity in the line-of-sight velocity distribution, and find values larger than 
the median of the \textsc{CHEX-MATE} population (0.57 versus 0.44).
As mentioned by \citet{Pizzuti_2025}, a more detailed dynamical analysis focussed
on \lens\,will be presented
by these authors.

While awaiting this analysis, we can already draw some conclusions based on the publicly available
spectroscopic data.

An offset between the velocity of the BCG and the mean velocity of the cluster members is commonly used as a diagnostic of a major cluster disturbance.
The BCG has a redshift of $z = 0.31665$ \citep{Driver_2022}, whereas the mean cluster redshift 
is $z = 0.31533$ \citep{Sereno_2025}.
Therefore, the BCG exhibits a line-of-sight velocity offset of $\sim$ 300\,km\,s$^{-1}$ 
relative to the cluster mean, suggesting that the system may not be fully dynamically relaxed. 
However, this alone is insufficient to confirm the presence of a major merger.
Such offset can be produced by several effects without invoking a major merger (wobbling of the BCG around the centre of the potential; substructure in the redshift distribution; projection effects; uncertainty in the cluster mean redshift).

Another relevant question for our study is whether we are able to disentangle \lens\, and
\lens b in velocity space and eventually estimate a dynamical mass for each component.

BCG$_2$ has a redshift of $z = 0.31708$ \citep{Proust_2015}. This translates into a
velocity difference equal to 98\,km\,s$^{-1}$ with respect to the BCG.
This is comparable to the typical uncertainty on the redshift measurement
\citep[80\,km\,s$^{-1}$,][]{Pizzuti_2025}.
Although this prevents us from resolving \lens\, and \lens b in velocity space, the small
offset suggests that the merger axis is likely oriented predominantly in the plane of the sky,
along the line connecting the BCG and BCG$_2$, \emph{i.e.} along the principal axis of the 
cluster. This is further supported by the morphology of the X-ray and radio emission contours.

\subsection{BCG-DM alignment}
We revisit the alignment between the position angles of the BCG and the dark matter halo, and examine its consistency with the merger scenario proposed here. At first glance, such an alignment might be interpreted as indicative of a relaxed, rather than a disturbed, dynamical state. However, this is not the case, as the observed alignment is fully consistent with the major merger scenario we advocate. Moreover, this alignment can be interpreted as further evidence that the system is being observed in a post-collisional phase of a major merger.

Cosmological hydrodynamical numerical simulations by \citet{Ragone_2020} show that major
mergers can disrupt the alignment.
Nevertheless, after some time (of order $\sim$Gyr) without further major perturbations,
the alignment is restored.
Moreover, mergers occurring along the principal axis of the cluster, as is likely the case here, tend to affect the alignment less than off-axis mergers.

On the observational side, \citet{Wittman_2019} studied the orientations of BCGs and their
host clusters in a sample of 22 clusters undergoing major mergers.
Interestingly, they selected clusters with properties similar to those of \lens-\lens b:
binary mergers observed after first pericentric passage, as inferred from their
radio properties, \emph{i.e.} typically $\sim$1\,Gyr after the collision.
Moreover, these systems exhibit unimodal line-of-sight velocity distributions, suggesting
the mergers occured predominantly in the plane of the sky.
Finally, the projected separations are typically of the order of $\sim$1\,Mpc.
They report an alignment consistent with that observed in the general cluster population, further supporting the interpretation that the alignment we find here is fully consistent with the major merger scenario we propose.

Altogether, these simulation and observational results strengthen the case for \lens\ being observed in the post-collisional phase of a major merger.

\subsection{A picture of the merger}

We propose a qualitative picture of the merger in \lens.

Our strong-lensing analysis reveals a significant mass concentration to the north-west, where 
an overdensity of bright galaxies is located.
This companion cluster, \lens b, is therefore the most likely candidate involved in the major merger,
perfectly aligned with the elongation of the X-ray and radio contours.

We are not able to derive a precise mass for \lens b at this stage and we note that our
mass estimates are indirect only. Nevertheless, we argue that 
\lens b is likely to have a cluster scale mass. When included in the strong lensing modelling as 
an external shear, it yields a shear strength of $\sim$0.13, a typical value at a projected distance 
of $\sim$200$\arcsec$ from the centre of a galaxy cluster\footnote{This is the order of 
magnitude of the shear measured at similar distances from the centre of Abell~1689,
\citep[see][]{mypaperIII}.}.
Alternatively, modelling \lens b as a singular isothermal sphere centred on BCG$_2$ results in a 
velocity dispersion of $750 \pm 150$\,km\,s$^{-1}$, which is characteristic of a galaxy cluster.
Considering the lower bound of 600\,km\,s$^{-1}$, and using the scaling relations from, 
\emph{e.g.}, \citet{Munari_2013}, this velocity dispersion can be associated with a 
halo with an expected global temperature of $\sim$ 2.5 keV, typical of a group-like structure 
hosting an X-ray–emitting gas component.

Finally, \citet{Krick_2007}
report the presence of diffuse ICL associated with \lens b, indicating past galaxy
interactions and a dark matter halo massive enough to trap stars released during these
interactions.

Therefore, \lens b is likely massive enough that, prior to the collision, it hosted an X-ray emitting
gas component.

The absence of an X-ray peak at the position of \lens b in the \textsc{XMM-Newton} data suggests 
that its gas has already been stripped, consistent with a post-merging 
phase. The X-ray tail to the South-East could then be interpreted as the result of 
interactions between the gas components of both clusters: while the dark matter and 
galaxies of \lens b are found to the North-West, its gas lags behind to the South-East, 
interacting with the gas of \lens.
Such X-ray emitting stripped gas mass clumps have been reported in other merging clusters,
for example MACS\,0717 \citep{Jauzac_2018} and Abell\,2744 \citep{Merten_2011,Jauzac_2016}.

Within this scenario, \lens\, is likely the more massive system, as indicated by the X-ray 
centroid being located close to its position. The relative amount of ICL associated with 
\lens\, and \lens b further supports this mass segregation, as does our SL analysis.

\section{Conclusions}

We have presented a new strong lensing and X-ray study of \lens\, using JWST, 
HST, \textsc{Chandra}, and \textsc{XMM-Newton} data. The unprecedented quality 
of the JWST imaging allowed us to more than double the number of
SL constraints and to identify conjugated sub-spots in several of the sixteen systems.
This leads to increasing the number of constraints and to quantify the improvements brought by the 
inclusion of sub-spots within certain multiple images.

Using these SL constraints, we construct a parametric mass model which explicitly incorporates the
\textsc{Chandra} data into a combined SL+X-ray fit, using the recently developed
methodology described in \citet{Beauchesne_2024}.
This is only the second application of this functionnality, the first being the study of
AS\,1063 by \citet{Beauchesne_2024,Beauchesne_2025a,Beauchesne_2025b}.
This allows us to disentangle the DM and the X-ray emitting components.
The DM component in \lens\, is found to be unimodal, spatially coincident with the
BCG, and characterized by a large core radius, 83$\pm$5\,kpc, consistent with values reported in other SL clusters. 

Adding the \textsc{XMM} data provides important clues, further indicating that \lens\, is undergoing 
a major merger with a North-Western 
companion, \lens b, now largely devoided of its hot gas. This scenario naturally explains 
the strong external shear, the X-ray and radio 
evidence of violent dynamics (X-ray tail and thermodynamic properties, radio relic, central radio halo), 
and the very high velocity dispersion.

This work demonstrates the value of fully combined SL+X-ray modelling in the 
JWST era, particularly for systems with complex assembly histories.

A next step to this work would be to observe \lens b with JWST in order to 
probe directly its mass distribution using possible SL signatures and/or weak lensing.
This would also allow to
test for any possible offset between its DM and stellar components. 
If DM is collisionless as proposed in the CDM scenario, the association between mass and 
light (stars) should be perfect, \emph{i.e.} the offset between the peaks of each component should
be consistent with 0 \citep{Roche_2024}. In contrast, if DM is self interacting, 
which is more likely to occur
during a cluster merger, the DM could experience a drag force while the stars would not, 
leading to a possibly measurable offset \citep{Valdarnini_2024,Sirks_2024}. 

In addition, the recently launched \textsc{xrism} satellite \citep{xrism}, with the
\textsc{Resolve} microcalorimeter \citep{Ishisaki_2022} can
measure gas velocities in the ICM, offering a promising avenue to better understand 
the geometry of the merger \citep[see, \emph{e.g.}][]{Heinrich_2025}.
\lens\, is also scheduled to be observed by \textit{Euclid}, enabling wide-field weak-lensing 
studies, in particular leading to a direct mass estimate of \lens b. Combined with deeper spectroscopy of cluster members, these observations will 
provide complementary constraints on the large-scale structure and the kinematic state 
of the merger, yielding a more complete picture of mass assembly in \lens.

Our mass model and associated products are made publicly available for download at the
\href{https://data.lam.fr/sl-cluster-atlas/}{Strong Lensing Cluster Atlas Data Base},
which is hosted at Laboratoire d'Astrophysique de
Marseille\footnote{https://data.lam.fr/sl-cluster-atlas/home.}.

\begin{acknowledgements}
This work is based in part on observations made with the NASA/ESA/CSA James Webb Space Telescope. The data were obtained from the Mikulski Archive for Space Telescopes at the Space Telescope Science Institute, which is operated by the Association of Universities for Research in Astronomy, Inc., under NASA contract NAS 5-03127 for JWST. These observations are associated with 
program 5594.
All our data products are available at MAST as a High Level Science Product
via \href{https://archive.stsci.edu/doi/resolve/resolve.html?doi=10.17909/z77s-5c44}{10.17909/z77s-5c44}.

We acknowledge the referee's careful and constructive reports that 
helped to significantly strengthen the manuscript and improve its clarity.
ML acknowledges discussions with Rapha\"el Gavazzi, Roser Pell\'o and Jessica Krick.
ML acknowledges the Centre National de la Recherche Scientifique (CNRS) and the
Centre National des Etudes Spatiale (CNES) for support.
BB acknowledges the Swiss National Science Foundation (SNSF) for supporting this work.
This work was performed using facilities offered by CeSAM (Centre de donnéeS Astrophysique de Marseille).
Centre de Calcul Intensif d’Aix-Marseille is acknowledged for granting access to its high performance computing resources.
MJ is supported and BB acknowledges partial support by the United Kingdom Research and Innovation (UKRI) Future Leaders Fellowship `Using Cosmic Beasts to uncover the Nature of Dark Matter' (grant number MR/X006069/1).
MB and FG acknowledge the financial contribution from the INAF GO grant 1.05.24.02.10 Extended Radio Emission in Galaxy Clusters at deep focus with MeerKAT.
MB acknowledges support from the ERC CoG $\vec{B}$ELOVED, GA N.101169773.
Support for program JWST-GO-5594 was provided by NASA through a grant from the Space Telescope Science Institute, which is operated by the Association of Universities for Research in Astronomy, Inc., under NASA contract NAS5-03127.

\end{acknowledgements}

\bibliographystyle{aa} 
\bibliography{draft}

\begin{appendix}

\section{Multiple images}
Table~\ref{multiple} list the multiple images.

\begin{table*}
\begin{center}
\begin{tabular}{cccccc}
\hline
\smallskip
Former ID  & New ID  & R.A. & Decl. & $z_{\rm spec}$ & $z_{\rm model}$ \\
\hline
S1 & 1.1& 344.70368& -34.798406 & 1.86 & \\
S2 & 1.2& 344.70045& -34.799264 & 1.867&  \\
S3 & 1.3& 344.69284& -34.805879 &   & \\
\hline
A1&2.1 & 344.70698&  -34.798195 & & \\
A2&2.2 & 344.69948&  -34.801443 & 1.869 & \\
A3&2.3 & 344.69552&  -34.805826 & & \\
A4&2.4 & 344.70178&  -34.803511 & & \\
A5&2.5 & 344.70135&  -34.802603 & & \\
\hline
B1& 3.1 & 344.70130 & -34.795580 & & $1.35 \pm 0.03$ \\ 
B2&3.2 & 344.69536 & -34.798629 & & \\ 
B3&3.3 & 344.69442 & -34.799193 & & \\ 
B4&3.4 & 344.69416 & -34.799334 & & \\ 
B5&3.5 & 344.69339 & -34.800246 & & \\
\hline
C1&4.1 & 344.70275 & -34.794065 & & \\
C2&4.2 & 344.69448 & -34.797526 & & \\
C3&4.3 & 344.69246 & -34.799303 & 2.854 & \\
\hline
D1&5.1 & 344.69710 & -34.804713 & & $ 1.38 \pm 0.04 $ \\ 
D2&5.2 & 344.69888 & -34.803751 & & \\ 
D3&5.3 & 344.69820 & -34.802791 & & \\ 
D4&5.4 & 344.69930 & -34.804344 & & \\ 
D5&5.5 & 344.70753 & -34.797873 & & \\     
\hline
E1&6.1 & 344.69487 & -34.806320  & 3.347 & \\
E2&6.2 & 344.70197 & -34.804498& & \\  
E3&6.3 & 344.70096 & -34.802696 & & \\
E4&6.4 & 344.69906 & -34.801483 & & \\
E5&6.5 & 344.70853 & -34.797417 & & \\
\hline
 & 7.1 &344.70371 &-34.808614 &  & $ 5.0 \pm 1.0 $\\
& 7.2 & 344.70106 &-34.809743 &  & \\
\hline
&8.1 &344.70389 & -34.807641 &  & $ 1.28 \pm 0.06 $ \\
&8.2 & 344.70234& -34.808337  &  & \\
\hline
&9.1 &344.71108 &-34.803280 &  & $ 10^*$ \\
&9.2 &344.70669 &-34.807969 &  & \\
\hline
&10.1 &344.71033 &-34.805106 &  & $ 2.39 \pm 0.14 $ \\
&10.2 &344.70649 &-34.808592 &  & \\
&10.3 &344.70216 &-34.810730 &  & \\
\hline
&11.1 &344.70424 &-34.801347 &  & $ 1.27 \pm 0.04$ \\
&11.2 &344.70351 &-34.802640 &  & \\
&11.3 &344.69490 &-34.806995 &  & \\
\hline
&12.1 &344.70269 &-34.800236 &  &  $1.18 \pm 0.03 $ \\
&12.2 &344.70193 &-34.800411 &  & \\
&12.3 &344.69447 &-34.806042 &  & \\
\hline
&13.1 &344.70197 &-34.796186 &  & $1.87 \pm 0.05 $\\
&13.2 &344.69788 &-34.797672 &  & \\
&13.3 &344.69230 &-34.802744  &  & \\
\hline
&14.1 &344.70217 &-34.805556 &  & $ 0.97 \pm 0.02$ \\
&14.2 &344.69995 &-34.806415 &  & \\
&14.3 &344.70675 & -34.801208 &  & \\
\hline
&15.1 &344.69601 &-34.801881 &  & $ 6.76 \pm 2.0 $ \\
&15.2 &344.69558 &-34.802297 &  & \\
\hline
&16.1 &344.70622 &-34.800816 &  & $ 1.24 \pm 0.04$ \\
&16.2 &344.70343 &-34.804425 &  & \\
&16.3 &344.69688 &-34.806907 &  & \\
\hline
\smallskip
\smallskip
\end{tabular}
\end{center}
\caption{Multiple images used in this work. 
For systems already identified prior to the JWST observations,
we list the IDs from previous works and the new IDs proposed in this study.
We report the spectroscopic redshift when available, as well as the 
estimated redshift constrained by the best fit mass model (with 1$\sigma$ error bars) presented in
Section~\ref{bestfit}. The estimated redshift for system 9 is stuck to 10, the upper bound of the prior.}
\label{multiple}
\end{table*}

\subsection{Sub-spots in multiply imaged systems}
\label{subspots_Appendix}

\subsubsection{Presentation}

We have been able to conjugate sub-spots in the following multiply imaged systems:
system 1 (12 sub systems), system 2 (4 sub systems), 
system 3 (2 sub systems), system 4 (6 sub systems), system 6 (2 sub systems), system 7 (4 sub systems)
and system 11 (2 sub systems).

We have been associating the most secure sub-spots on each given image in order not to introduce
misidentification.
Sometimes we have not been able to conjugate sub-spots for all images of a given system.
For example, we have not been able to reliably define image 2.3.5, hence this image is not mentioned
in Table~\ref{subspots_1}.
In defining the sub-spots, we have set that the former image $(i.j)$ becomes $(i.1.j)$.
For example, image 2.1 becomes 2.1.1 and image 2.2 becomes 2.1.2.

All images are listed in Table~\ref{subspots_1} and \ref{subspots_2}.
We show the identifications proposed on Fig.~\ref{fig_subspots_sys1}
and Fig.~\ref{fig_subspots}.
Color images were produced with SAOImage ds9, using JWST/NIRCam F322W, F150W, and HST/ACS F814W in the red, green, and blue channels, respectively.

\subsubsection{What Do Sub-Spots Contribute?}

\begin{figure*}
\begin{center}
\includegraphics[scale=0.5,angle=0.0]{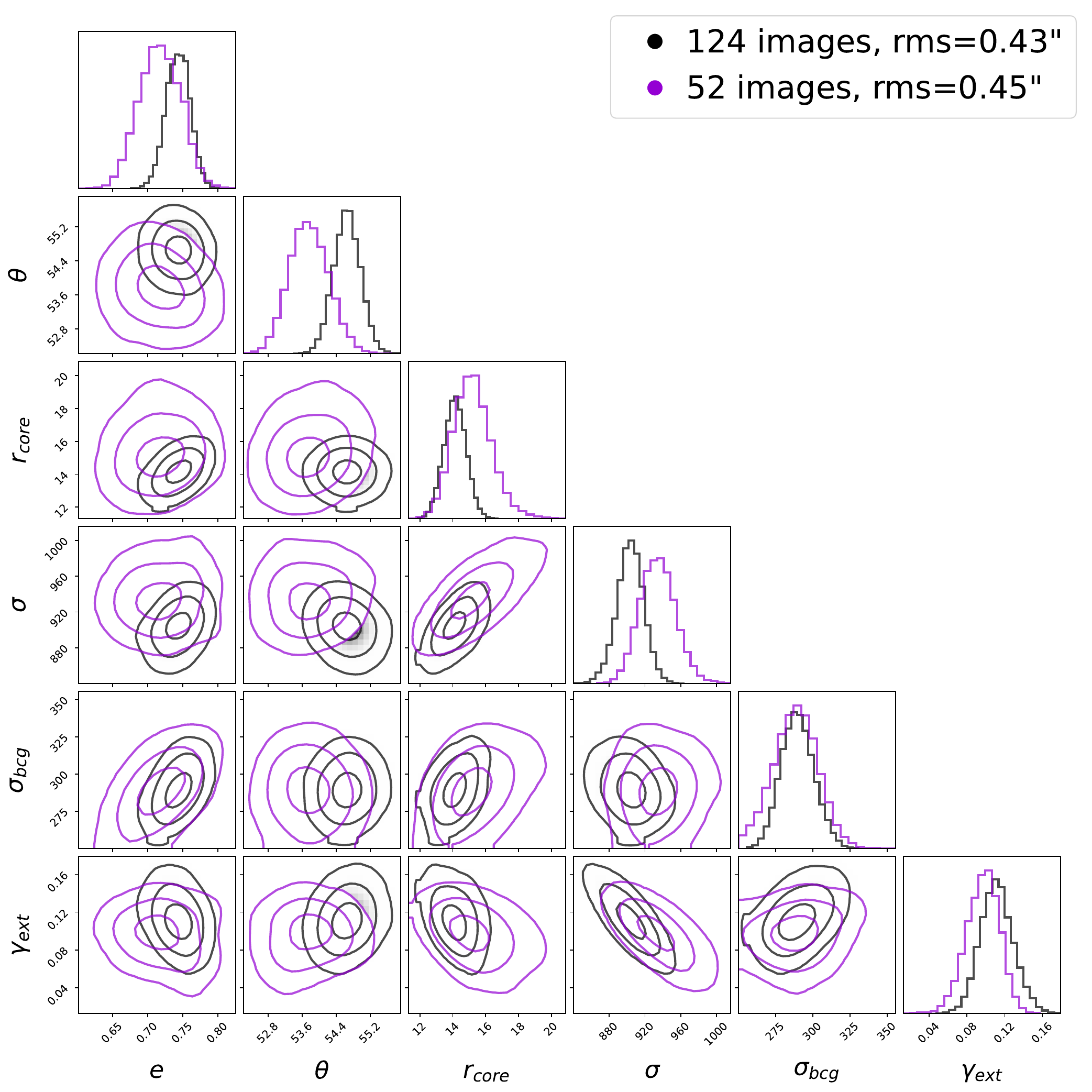}
\caption{Posterior distributions of the main clump parameters from the
SL+X-ray fits performed using either 52 or 124 images.}
\label{52_124}
\end{center}
\end{figure*}

\begin{figure*}
\begin{center}
\includegraphics[scale=0.28,angle=0.0]{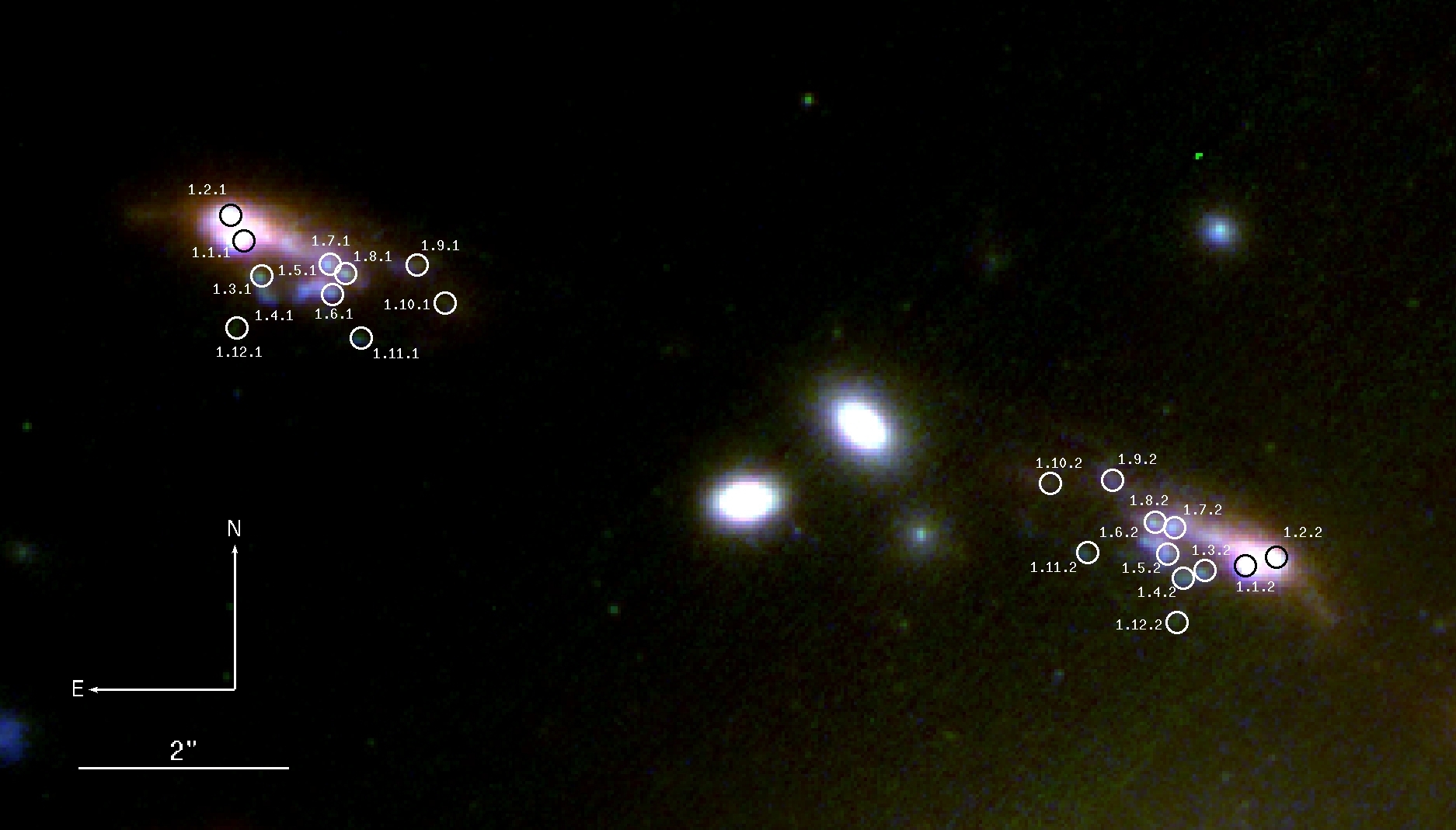}
\includegraphics[scale=0.5,angle=0.0]{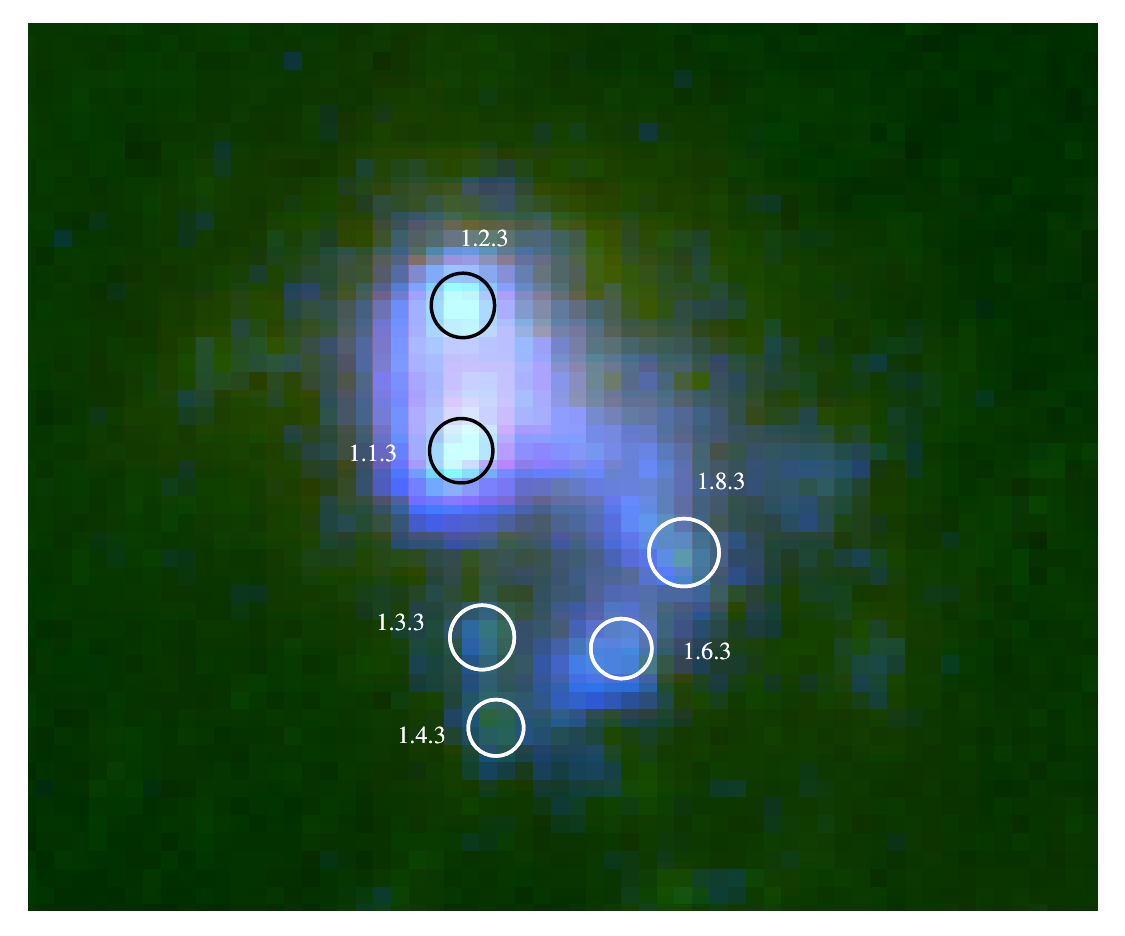}
\caption{Identification of the sub-spots in system 1.
}
\label{fig_subspots_sys1}
\end{center}
\end{figure*}

\begin{figure*}
\begin{center}
\includegraphics[scale=0.47,angle=0.0]{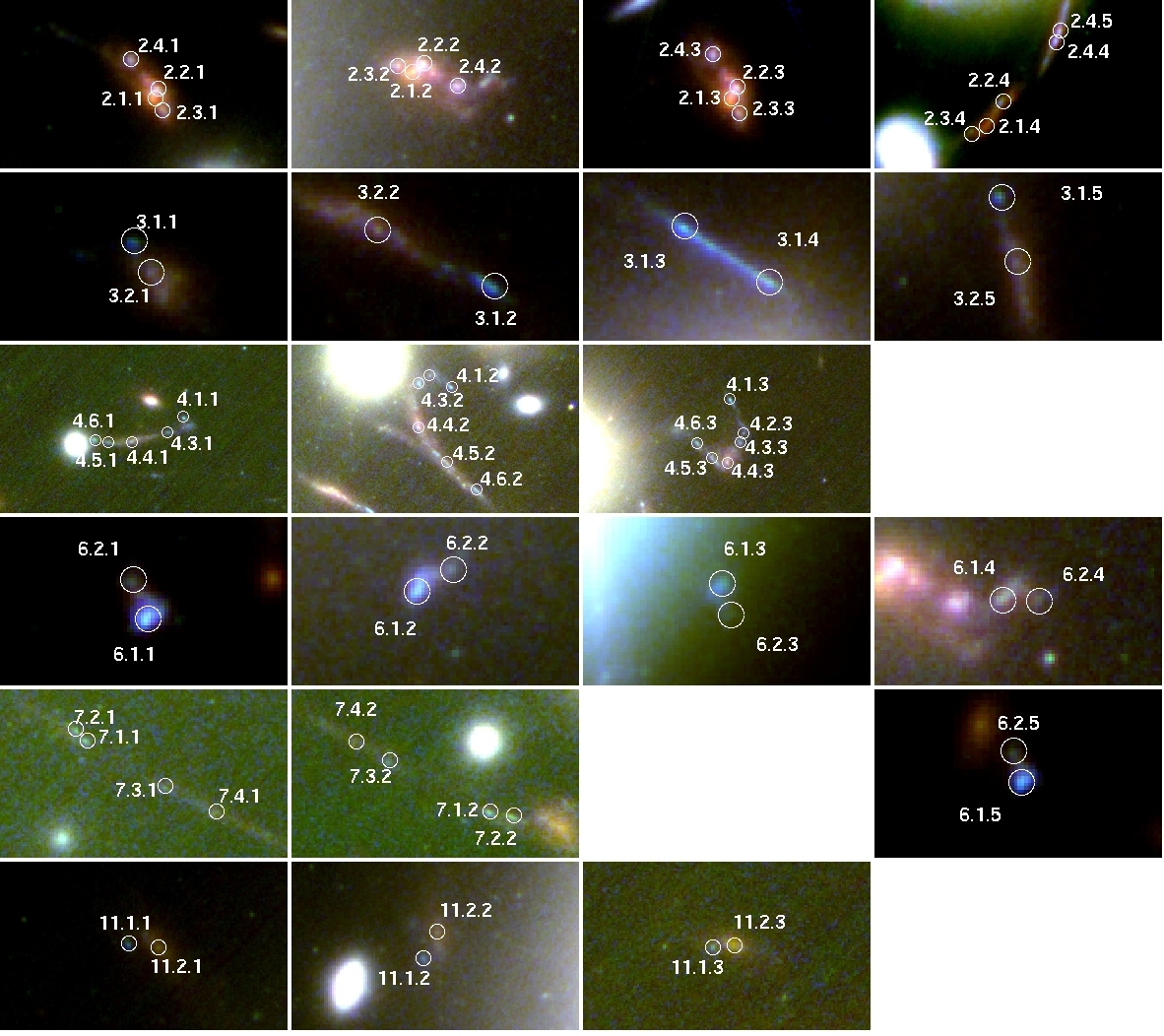}
\caption{Identification of the sub-spots in systems 2,3,4,6,7 and 11.
}
\label{fig_subspots}
\end{center}
\end{figure*}

We aim to quantify the improvement in constraints brought by the inclusion of sub-spots within 
certain multiple images (Section~2.3). In Fig.~\ref{52_124}, we present the posterior distributions of the main clump 
parameters derived from the SL+X-ray fits performed using either 52 or 124 images.
The PDFs are generally narrower when using 124 images. In particular, the PDFs for the velocity 
dispersion of the DM clump and the strength of the external shear are more sharply peaked and do not 
exhibit a plateau, as observed in the case of 52 images.

\begin{table}
\begin{center}
\begin{tabular}{ccccc}
\hline
\smallskip
 ID  & R.A. & Decl. & $z_{\rm spec}$ & $z_{\rm model}$ \\
\hline

1.1.1 &  344.70368  & -34.798406 & 1.867 & \\ 
1.1.2 &  344.70045  & -34.799264 & &\\ 
1.1.3  & 344.69284  & -34.805879 & &\\ 
1.2.1  & 344.70372  & -34.798335 &1.867 & \\ 
1.2.2  & 344.70035  & -34.799241 & &\\ 
1.2.3  & 344.69283  & -34.805800 & & \\ 
1.3.1  & 344.70362  & -34.798497 & 1.867 & \\ 
1.3.2  & 344.70058  & -34.799278 & & \\ 
1.3.3  & 344.69282  & -34.805969 & & \\ 
1.4.1  & 344.70360  & -34.798550 & 1.867 & \\
1.4.2  & 344.70065  & -34.799297 & & \\ 
1.4.3  & 344.69282  & -34.806008 & & \\ 
1.5.1  & 344.70347  & -34.798536 & 1.867 & \\ 
1.5.2  & 344.70070  & -34.799233 & & \\ 
1.6.1  & 344.70339  & -34.798547 & 1.867 & \\  
1.6.2  & 344.70077  & -34.799197 & & \\
1.6.3  & 344.69274  & -34.805975& & \\ 
1.7.1  & 344.70340  & -34.798467 & 1.867 & \\
1.7.2  & 344.70068  & -34.799164& & \\ 
1.8.1  & 344.70335  & -34.798492 &1.867 & \\
1.8.2  & 344.70074  & -34.799150& & \\ 
1.8.3  & 344.69271  & -34.805922& & \\ 
1.9.1  & 344.70312  & -34.798469 & 1.867 & \\
1.9.2  & 344.70087  & -34.799033& & \\ 
1.10.1  & 344.70303  & -34.798569 &1.867 & \\ 
1.10.2  & 344.70108  & -34.799047& & \\ 
1.11.1  & 344.70330  & -34.798661 & 1.867 & \\ 
1.11.2  & 344.70096  & -34.799231& & \\ 
1.12.1  & 344.70370  & -34.798636& 1.867 & \\ 
1.12.2  & 344.70067  & -34.799414& & \\
\hline
2.1.1 & 344.70698 & -34.798195 & 1.869 & \\
2.1.2 & 344.69948 & -34.801443 & & \\ 
2.1.3 & 344.69552 & -34.805826 & & \\ 
2.1.4 & 344.70178 & -34.803511 & & \\ 
2.1.5 & 344.70135 & -34.802603 & & \\ 
2.2.1 & 344.70696 & -34.79815 &  1.869 & \\
2.2.2 & 344.69941 & -34.801408 & & \\ 
2.2.3 & 344.69548 & -34.805772 & & \\ 
2.2.4 & 344.70169 & -34.803403 & & \\ 
2.2.5 & 344.70136 & -34.802692 & & \\ 
2.3.1 & 344.70694 & -34.798244 & 1.869 & \\ 
2.3.2 & 344.69955 & -34.801419 & & \\ 
2.3.3 & 344.69548 & -34.805892 & & \\ 
2.3.4 & 344.70186 & -34.803544 & & \\ 
2.4.1 & 344.70711 & -34.798019 & 1.869 & \\
2.4.2 & 344.69923 & -34.801506 & & \\ 
2.4.3 & 344.69561 & -34.805625 & & \\
2.4.4 & 344.7014  & -34.803138 &  & \\
2.4.5 & 344.70138 &  -34.803089 &   & \\
\hline
3.1.1  & 344.70130 &  -34.795580 &  &  $1.35 \pm 0.03 $ \\  
3.1.2  & 344.69536  & -34.798629 &  &  \\ 
3.1.3  & 344.69442  & -34.799193 &  &  \\ 
3.1.4  & 344.69416  & -34.799334 &  &  \\ 
3.1.5  & 344.69339  & -34.800246 &  &  \\ 
3.2.1  & 344.70125  & -34.795662 &  &  $1.29 \pm 0.03 $  \\
3.2.2  & 344.69572  & -34.798487 &  &  \\ 
3.2.5 &  344.69334  & -34.800412 &  &  \\
\hline
\smallskip
\smallskip
\end{tabular}
\end{center}
\caption{Systems for which we propose sub-systems: systems 1, 2 and 3.
We report the
estimated redshift constrained by the mass model (with 1$\sigma$ error bars) for each sub-system of system 3, 
in order to check that they are, as expected, in agreement.}
\label{subspots_1}
\end{table}

\begin{table}
\begin{center}
\begin{tabular}{ccccc}
\hline
\smallskip
 ID  & R.A. & Decl. & $z_{\rm spec}$ & $z_{\rm model}$ \\
\hline
4.1.1 & 344.70275 & -34.794068&  2.854 & \\ 
4.1.2 & 344.69448 & -34.797526&  & \\ 
4.1.3 & 344.69245 & -34.799311&  & \\ 
4.2.2 & 344.69473 & -34.797422&  2.854  & \\
4.2.3 & 344.69230 & -34.799617&  & \\ 
4.3.1 & 344.70292 & -34.794202&  2.854  & \\
4.3.2 & 344.69484 & -34.797489&  & \\ 
4.3.3 & 344.69234 & -34.799700&  & \\ 
4.4.1 & 344.70331 & -34.794289&  2.854 & \\ 
4.4.2 & 344.69485 & -34.797883&  & \\ 
4.4.3 & 344.69248 & -34.799878&  & \\ 
4.5.1 & 344.70356 & -34.794292&  2.854 & \\ 
4.5.2 & 344.69454 & -34.798189&  & \\ 
4.5.3 & 344.69264 & -34.799837&  & \\ 
4.6.1 & 344.70370 & -34.794272&  2.854 & \\
4.6.2 & 344.69422 & -34.798438& & \\ 
4.6.3 & 344.69281 & -34.799703& & \\
\hline
6.1.1 & 344.69487&  -34.806325 & 3.347 & \\ 
6.1.2 & 344.70198&  -34.804493 & & \\ 
6.1.3 & 344.70097&  -34.802690 & & \\ 
6.1.4 & 344.69909&  -34.801494 & & \\ 
6.1.5 & 344.70854&  -34.797411 & & \\ 
6.2.1 & 344.69491&  -34.80623 & 3.347 & \\ 
6.2.2 & 344.70188&  -34.804442 & & \\ 
6.2.3 & 344.70094&  -34.80277 & & \\ 
6.2.4 & 344.69898&  -34.801495 & & \\ 
6.2.5 & 344.70857&  -34.797338& & \\ 
\hline
7.1.1 & 344.70371 & -34.808614 & & $ 5.0 \pm 1.0 $ \\ 
7.1.2 & 344.70106 & -34.809743 & & \\ 
7.2.1 & 344.70377 & -34.808564 & & $ 6.1 \pm 0.8 $ \\ 
7.2.2 & 344.70094 & -34.809769 & & \\ 
7.3.1 & 344.70328 & -34.808819 & & $ 6.5 \pm 0.8 $\\ 
7.3.2 & 344.70161 & -34.809522 & & \\ 
7.4.1 & 344.70302 & -34.808933 & & $ 5.7 \pm 0.8 $ \\ 
7.4.2 & 344.70179 & -34.809442 & & \\ 
\hline
11.1.1 & 344.70424&  -34.801347 & & $ 1.27 \pm 0.04 $ \\ 
11.1.2 & 344.70351&  -34.802640 & & \\ 
11.1.3 & 344.69490&  -34.806995 & & \\ 
11.2.1 & 344.70408&  -34.801364 & & $1.28 \pm 0.04 $ \\ 
11.2.2 & 344.70344&  -34.802514 & & \\ 
11.2.3 & 344.69480&  -34.806983 & & \\ 
\hline
\smallskip
\smallskip
\end{tabular}
\end{center}
\caption{Systems for which we do propose sub-systems: systems 4, 6, 7 and 11.
We report the
estimated redshift constrained by the mass model (with 1$\sigma$ error bars) for each sub-system when no spectroscopic redshift is available,
in order to check that they are, as expected, in agreement.}
\label{subspots_2}
\end{table}

\section{The (\textsc{rate}, Nb) Test}
\label{RateNbAppendix}

\begin{figure*}
\begin{center}
\includegraphics[scale=0.5,angle=0.0]{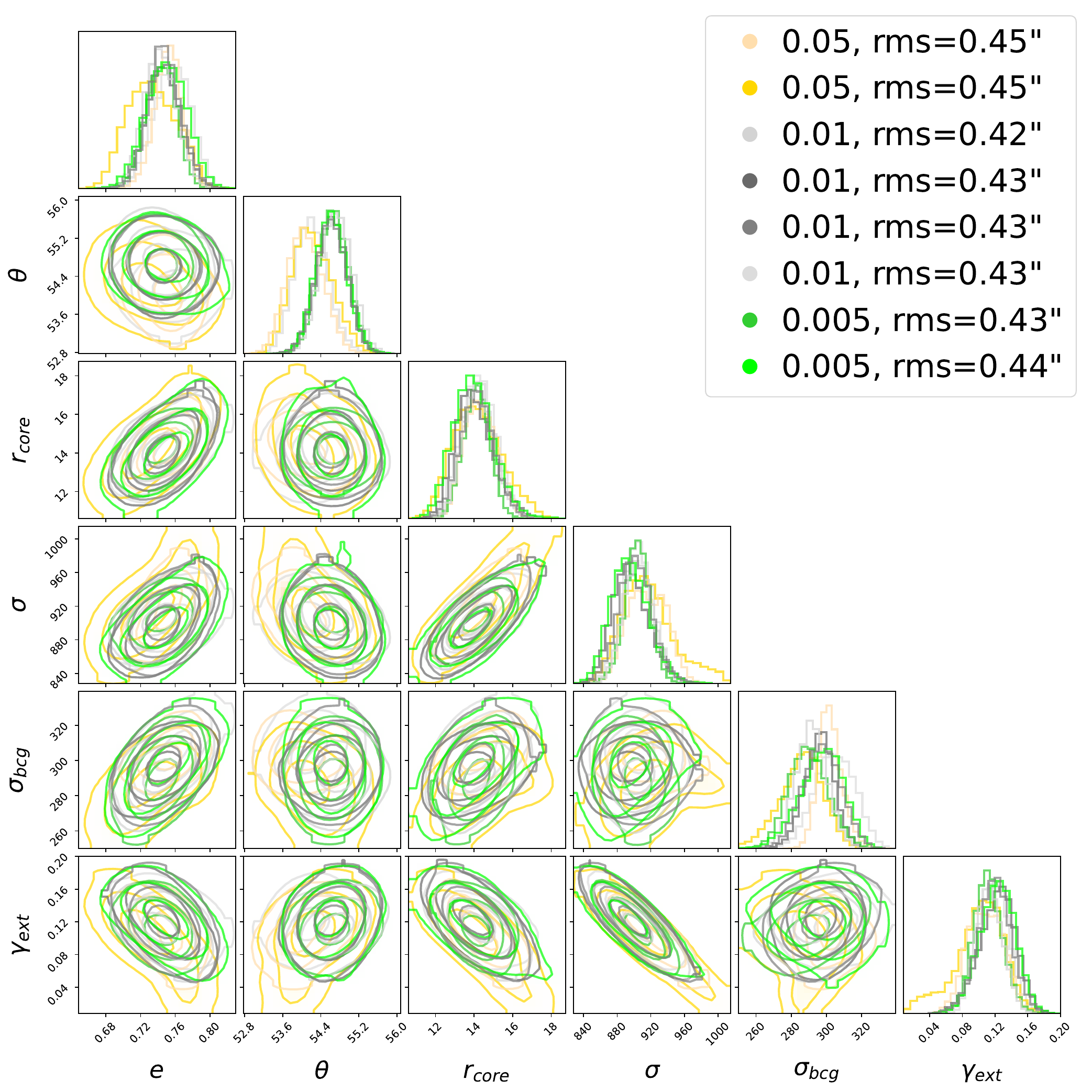}
\caption{Results of the (\textsc{Rate}, \textsc{Nb}) Test. We show the 
posterior distributions of the main cluster parameters from the SL+X-ray fits performed 
for the different (\textsc{Rate}, \textsc{Nb}) values listed in Table~\ref{rateNb} and
indicated on the upper right inset of the figure.}
\label{fig_rateNb}
\end{center}
\end{figure*}

We perform the (\textsc{Rate}, \textsc{Nb})
test, which has been found to be useful for assessing whether the parameters
describing the mass distribution have actually converged.
Two key parameters matter when it comes to the convergence of the \textsc{Lenstool} MCMC sampler
\citep{jullo07}: the \textsc{rate} parameter, associated to the burnin phase,
and the number of iterations (\textsc{Nb}), associated to the
sampling phase.
The smaller the rate, the more the sampler will move slowly to the high-likelihood areas and will be less prone to miss a mode of the posterior.
The larger \textsc{Nb}, the larger the number of iterations of the MCMC chains, the best the
parameter space can be sampled.
We refer the reader to \citet{Limousin_2025} for more details.
In brief, we decrease the value of \textsc{Rate} and increase \textsc{Nb} to evaluate their 
influence on the
RMS and on the parameters of the mass clumps describing \lens\, in the context of the combined
SL+X-ray analysis.
Convergence is considered to be attained when the values of these parameters no longer 
influence the resulting RMS or the parameters of the mass clumps. The results are 
presented in Table~\ref{rateNb} and Fig.~\ref{fig_rateNb}.
We find that a \textsc{Rate} value of 0.01 is sufficiently small: the PDFs corresponding to the 
different runs with \textsc{Rate} = 0.01 are in good agreement with one another.
In contrast, the PDFs obtained with \textsc{Rate} = 0.05 are broader, suggesting
that this value is not small enough.
Moreover, when lowering the \textsc{Rate} to 0.005, the resulting PDFs remain 
consistent with those obtained using \textsc{Rate} = 0.01.
Regarding \textsc{Nb}, we find no significant differences between runs with values of 2000 or 
4000, suggesting that when the \textsc{Rate} is sufficiently small, \textsc{Nb} becomes less 
critical, as also found in \citet{Limousin_0416} for MACS\,J0416.

\begin{table}
\begin{center}
\begin{tabular}{ccc}
\hline \\*[-1mm]
\textsc{rate} & \textsc{Nb} & RMS ($\arcsec$) \\
\hline \\*[-1mm]
0.05 & 2000 & 0.45 \\
\hline \\*[-1mm]
0.05 & 4000 & 0.45 \\
\hline \\*[-1mm]
0.01 & 2000 & 0.42 \\
\hline \\*[-1mm]
0.01 & 2000 & 0.43 \\
\hline \\*[-1mm]
0.01 & 3000 & 0.43 \\
\hline \\*[-1mm]
0.01 & 4000 & 0.44 \\
\hline \\*[-1mm]
0.005 & 2000 & 0.43 \\
\hline \\*[-1mm]
0.005 & 2000 & 0.44 \\
\hline \\*[-1mm]
\smallskip
\end{tabular}
\end{center}
\caption{RMS values obtained for various combinations of \textsc{Rate} and \textsc{Nb}.}
\label{rateNb}
\end{table}

\end{appendix}

\end{document}